\documentclass[10pt, journal, twocolumn]{IEEEtran}
\usepackage{amsmath}

\usepackage{amssymb,amsthm}
\usepackage{bm}
\usepackage{color}
\usepackage{graphicx}
\usepackage{empheq}
\usepackage[linesnumbered,ruled,lined]{algorithm2e}
\usepackage[outdir=./]{epstopdf}
\usepackage{caption}
\usepackage{lipsum}
\usepackage{cite}
\usepackage{multirow}
\usepackage{subcaption}
\usepackage{fancyhdr}
\SetKwInput{Input}{input}
\SetKwInput{Output}{output}
\newtheorem{definition}{Definition}
\usepackage{array}
\newcolumntype{L}[1]{>{\raggedright\let\newline\\\arraybackslash\hspace{0pt}}m{#1}}
\newcolumntype{C}[1]{>{\centering\let\newline\\\arraybackslash\hspace{0pt}}m{#1}}
\newcolumntype{R}[1]{>{\raggedleft\let\newline\\\arraybackslash\hspace{0pt}}m{#1}}

\newtheorem{proposition}{Proposition}
\theoremstyle{remark}
\newtheorem{remark}{Remark}
\IEEEoverridecommandlockouts

\begin{document}

	\title{Beyond Dirty Paper Coding for Multi-Antenna Broadcast Channel with Partial CSIT: \\A Rate-Splitting Approach
   }

		\author{
			
			\IEEEauthorblockN{Yijie Mao, \IEEEmembership{Member, IEEE}, Bruno Clerckx, \IEEEmembership{Senior Member, IEEE
				}  }

			\thanks{
				Y. Mao and B. Clerckx are with Imperial College London, London SW7 2AZ, UK (email: y.mao16@imperial.ac.uk; b.clerckx@imperial.ac.uk).\newline
				This work has been partially supported by the U.K. Engineering and Physical Sciences Research Council (EPSRC) under grant EP/N015312/1, EP/R511547/1.}

			}

\maketitle

\thispagestyle{empty}
\pagestyle{empty}
\begin{abstract}
Imperfect Channel State Information at the Transmitter (CSIT) is inevitable in modern wireless communication networks, and results in severe multi-user interference in multi-antenna Broadcast Channel (BC). While the capacity of  multi-antenna (Gaussian) BC with perfect CSIT is known and achieved by Dirty Paper Coding (DPC), the capacity and the capacity-achieving strategy of  multi-antenna BC with imperfect CSIT remain unknown.  Conventional approaches therefore rely on applying communication strategies designed for perfect CSIT to the imperfect CSIT setting. In this work, we break this conventional routine and make two major contributions. First, we show that linearly precoded Rate-Splitting (RS), relying on the split of messages into common and private parts and linear precoding at the transmitter, and successive interference cancellation at the receivers, can achieve larger rate region than DPC in multi-antenna BC with partial CSIT. 
Second, we propose a novel  scheme, denoted as Dirty Paper Coded Rate-Splitting (DPCRS), that relies on RS to split the user messages into common and private parts, and DPC to encode the private parts. We show that the  rate region of DPCRS in Multiple-Input Single-Output (MISO) BC with partial CSIT is enlarged beyond that of conventional DPC and that of linearly precoded RS. Gaining benefits from the capability of RS to partially decode the interference and partially treat interference as noise, DPCRS is less sensitive to CSIT inaccuracies, networks loads and user deployments compared with DPC and other existing transmission strategies. 
\end{abstract}

\section{Introduction}
\par Current wireless communication networks rely increasingly on multi-antenna/Multiple-Input Multiple Output (MIMO) processing to boost rate performance and manage interference. Although appealing in their concept, multi-antenna networks are nevertheless hampered by several practical factors. Among these, the acquisition of accurate Channel State Information (CSI) knowledge at the Transmitter (CSIT) is a major challenge. The availability of accurate CSIT is crucial for downlink multi-user multi-antenna wireless networks. The beamforming and interference management performance heavily depends on the channel estimation accuracy. Unfortunately, pilot reuse tends to impair channel estimation with inter-cell interference in Time Division Duplex (TDD) and a significant feedback overhead is required to guarantee sufficient feedback accuracy in Frequency Division Duplex (FDD) due to the potentially large number of antennas. Delay, mobility, Radio Frequency (RF) impairments (e.g. phase noise) and inaccurate calibrations of RF chains also contribute to making the CSIT inaccurate. Moreover, CSI may be known only for a subset of  links in the network, may be estimated only at the subband level (and not for each subcarrier) and may not be known instantaneously but only statistically. This CSIT inaccuracy results in a multi-user interference problem that is the primary bottleneck of MIMO wireless networks. 
As an illustration of the severity of the problem, in 4G Long-Term Evolution (LTE)-Advanced, the CSIT inaccuracy leads to significant losses of Multi-User MIMO (MU-MIMO) of at least 30\% in terms of cell average throughput, and 42\% in terms of cell edge throughput \cite{clerckx2013mimo}. Similarly Coordinated Multi-Point (CoMP) transmission based on coordinated scheduling and beamforming across a full network leads to performance even worse than single-cell processing because of the inaccurate CSIT in the presence of subband-based feedback in 4G LTE-Advanced \cite{clerckx2013mimo}.

\par
\begin{figure}
	\centering
	\begin{subfigure}[b]{0.34\textwidth}
		\centering
		\includegraphics[width=\textwidth]{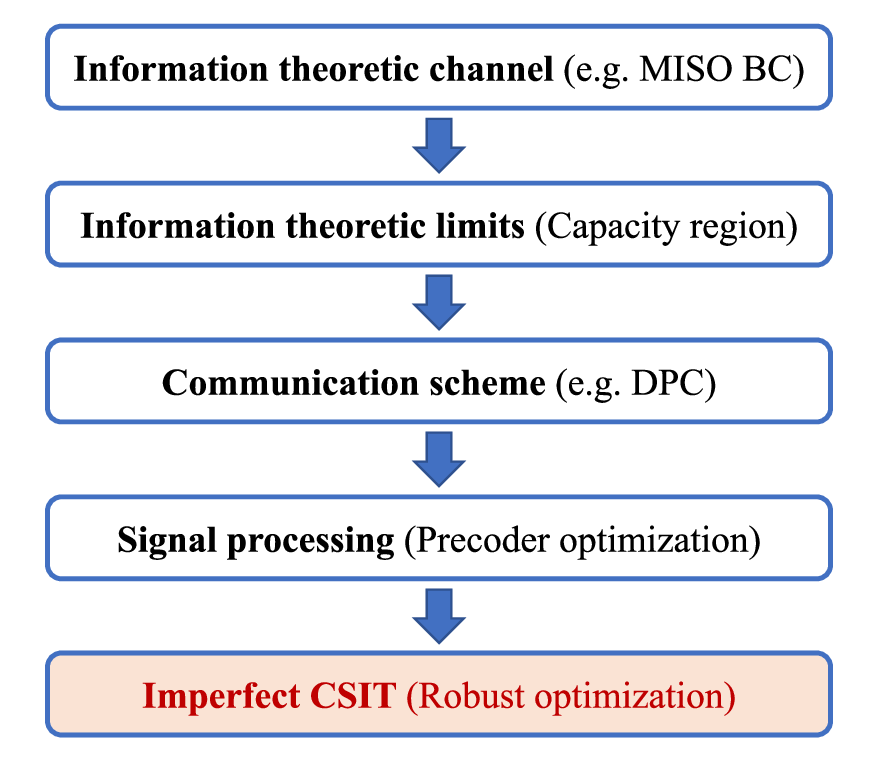}
		\caption{Conventional approach}
	\end{subfigure}
	~\\
	\begin{subfigure}[b]{0.44\textwidth}
		\centering
		\includegraphics[width=\textwidth]{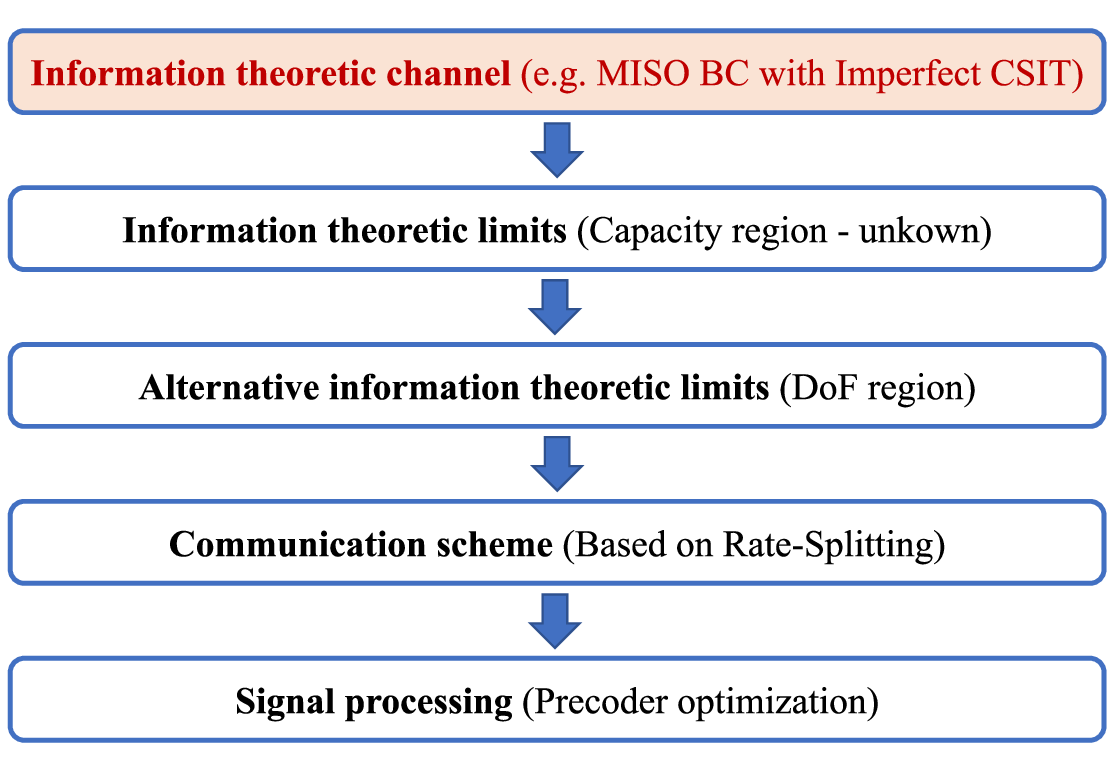}
		\caption{Approach motivated by CSIT}
	\end{subfigure}
	\caption{Conceptual flow of conventional  approach and partial CSIT-motivated approach for rate-splitting in MISO BC.}
	\label{fig: conceptualFlow}
\end{figure}

 Looking backward, the problem has been to strive to apply techniques designed for perfect CSIT to scenarios with partial CSIT \cite{RSintro16bruno}. Indeed, multi-antenna in 4G and 5G have been fundamentally motivated by the assumption of perfect CSIT and their performance assessed in the presence of partial CSIT. This is reflected by the conventional approach used in the past 20 years that consists in identifying a communication theoretic channel, e.g. Multiple-Input Single-Output (MISO) Broadcast Channel (BC), characterize its fundamental limits, e.g. capacity region, identify the capacity-achieving strategy, e.g. Dirty Paper Coding (DPC), simplify the strategy, e.g. using linear precoding, and then incorporate partial CSIT and design robust precoders. This leads to the classical linear precoding framework where any residual interference is treated as noise at the receivers. This conventional approach is further illustrated in Fig. \ref{fig: conceptualFlow}(a).    

\par While the ability to provide highly accurate and up-to-date CSIT remains questionable, considerable effort has been devoted to improving the performance of those strategies in the presence of CSIT uncertainties. Unfortunately, such approaches have been shown partially disappointing (e.g. CoMP in 4G as discussed above) and it is conjectured that following the same path will increase the gap between theory and practice as the density of antennas increases. The caveat of this conventional approach is that the underlying communication strategies were motivated by perfect CSIT, and partial CSIT is brought into the picture only at the end of the design. The rationale why such a conventional approach has been extensively used is that, while the capacity of the multi-antenna (Gaussian) BC with perfect CSIT is known and achieved by DPC \cite{ZFandZFDPC2003,capacityRegion2006HW}, the capacity and the capacity-achieving strategy of the multi-antenna BC with imperfect CSIT remain unknown.  However, such conventional approach may come with non-negligible performance loss. It has been shown in \cite{DPC2012NanLee, yang2005impact}  that DPC is  very sensitive to imperfect CSIT. As CSIT quality decreases, DPC region becomes much smaller. Below a certain threshold of CSIT quality, DPC region would be outperformed by Beamforming with Joing Decoding (BF--JD) strategy proposed in \cite{DPC2012NanLee} or even time-sharing \cite{yang2005impact} under the assumption of perfect  Channel State Information at the Receivers (CSIR).

\par In this paper, we consider another approach and wonder whether it would be wiser to design MIMO wireless networks from scratch accounting for partial CSIT and its resulting multi-user interference \cite{RSintro16bruno}. The fundamental question and first motivation for this paper is \textit{can we design and optimize transmission strategies for multi-user multi-antenna communications under the assumption of partial CSIT?} Interestingly, new communication and information theoretic understanding of the fundamental role of partial CSIT on the performance of MIMO wireless networks has appeared. It is now known that to benefit from partial CSIT and tackle the multi-user interference, the transmitter should take a \textit{Rate-Splitting (RS) approach} that splits the messages into a common and a private parts, encodes the commmon parts into a common stream, and private parts into private streams and superposes in a non-orthogonal manner the common stream on top of all private streams \cite{RSintro16bruno}. The {common stream} is drawn from a codebook shared by all receivers and is intended to one but is decodable by all receivers, while the {private streams} are to be decoded by their corresponding receivers only. Such approach is optimal from an information theoretic perspective (Degrees-of-Freedom, DoF) in a $K$-user MISO BC with partial CSIT \cite{AG2015,RS2016hamdi,enrico2017bruno,hamdi2019spawc}, and brings partial CSIT early on in the picture as illustrated in Fig. \ref{fig: conceptualFlow}(b).

\par This proposed approach contrasts with the conventional approach (as used in 4G and 5G) that is entirely designed based on private streams transmission. Importantly, the proposed RS approach is a more general framework that boils down to conventional  precoding whenever no power is allocated to the common stream. That has for consequence that RS-based approaches achieve equal or better performance compared to conventional precoding. Over the past few years, the benefits of  RS-based network design have been shown in the literature of MIMO wireless networks.  The dawn of RS in multi-antenna BC  is from an information theoretic analysis. RS is shown   to achieve the optimal sum DoF  \cite{RS2016hamdi} and  further proved to achieve the entire DoF region \cite{enrico2017bruno,hamdi2019spawc} of the $K$-user underloaded MISO BC with partial CSIT. The DoF benefits of RS are also studied in the underloaded BC with   multiple transmitters \cite{chenxi2017brunotopology} and multi-antenna receivers \cite{chenxi2017bruno} in partial CSIT. In the overloaded MISO BC, RS has been shown to achieve the optimal DoF region with heterogeneous CSIT qualities by superimposing degraded symbols on top of linearly precoded RS symbols.  The merits of RS discovered from  DoF analysis motivate recent studies   of precoder design for RS at finite Signal-to-Noise Ratio (SNR)  with both perfect and partial CSIT.  Specifically, RS linear precoders have been designed in the conventional MISO BC  for  sum rate maximization with partial CSIT  \cite{RS2016hamdi} and perfect CSIT \cite{mao2017rate},  max-min fair transmission with partial CSIT  \cite{hamdi2016robust}, energy efficiency maximization with perfect CSIT \cite{mao2018EE}, transmit power control with partial CSIT \cite{Medra2018SPAWC} and minimizing the mean square error with finite feedback \cite{Lu2018MMSERS}. Moreover, it has been  shown in \cite{shengYang2018ITW} that the entire capacity region of two-user MISO BC can be achieved by RS with Minimum Mean Square Error (MMSE) precoding within a constant gap.
 Besides linearly percoded RS, precoder design of RS with non-linear Tomlinson-Harashima Precoding (THP) in MISO BC has been studied in \cite{Flores2018ISWCS}. Though THP technique does not achieve the performance of DPC, it is less complex and considered as a practical implementation of DPC. The precoders of RS have also been designed in other variants of MISO BC, such as multi-group multicast \cite{hamdi2017bruno}, massive MIMO \cite{Minbo2016MassiveMIMO}, millimeter-wave systems \cite{minbo2017mmWave}, MISO BC with hardware impairments \cite{AP2017bruno}, CoMP joint transmission \cite{mao2018networkmimo}, Cloud Radio Access Networks (C-RAN) \cite{alaa2019IEEEaccess}, Simultaneous Wireless Information and Power Transfer (SWIPT) \cite{mao2019swipt}, Non-Orthogonal Unicast and Multicast  (NOUM) transmission \cite{mao2019TCOM}, cooperative RS in MISO BC with user relaying \cite{jian2019CRS,mao2019maxmin}. 
The capability of 1-layer RS  discovered in the literature to partially decode the interference and partially treat residual interference as noise makes RS the fundamental building block for a more general and powerful transmission framework for  downlink BC, namely, \textit{Rate-Splitting Multiple Access (RSMA)}  \cite{mao2017rate}. 
RSMA uses linearly precoded RS at the transmitter to split each user message into multiple common messages and a private message. The common messages are recombined and  encoded into the common streams  for the intended users. Successive Interference Cancellation (SIC)  is required  at each user to  sequentially decode the intended common streams. 
Such linearly precoded generalized RSMA has been demonstrated to be a powerful framework to bridge and generalize Space Division Multiple Access (SDMA) and Non-Orthogonal Multiple Access (NOMA), and further boost  system spectral and energy efficiencies for downlink MISO BC with both perfect and partial CSIT. 
As a summary, the developed framework based on RS is not only optimum from an information theoretic (DoF) perspective, it also provides significant performance benefits over the conventional precoding strategies.

\par Building upon the progress in the RS literature for multi-antenna BC, this paper studies RS and DPC in MISO BC with partial CSIT, and makes two major and novel contributions:

\par \textit{First}, this paper shows that linearly precoded RS outperforms DPC in MISO BC with partial CSIT. The performance benefits come from the inherent robustness of RS to partial CSIT. This is the first paper to explicitly make this observation. This is in sharp contrast with the perfect CSIT case, where DPC is known to be capacity achieving \cite{ZFandZFDPC2003} and outperform linearly precoded RS \cite{mao2017rate}. This has major implications for practical communication system designs. On one extreme, DPC can be seen as a full transmit-side interference cancellation strategy. On the other extreme, power-domain NOMA based on Superposition Coding (SC) and SIC can be seen as a full receive-side interference cancellation strategy. Power-domain NOMA, however, wastes the multiplexing gain of  MISO BC and results in poor performance, as explained in \cite{mao2017rate,bruno2019wcl}. In between, stands RS that can be seen as a smart combination of transmit-side and receive-side interference cancellation strategy where the contribution of the common stream is adjusted according to the level of interference that needs to be canceled by the receiver. What this paper shows is that, in practical deployments subject to partial CSIT, an RS strategy enabling a mix of transmit-side and receive-side interference cancellation outperforms a full transmit-side interference cancellation strategy such as DPC. This further demonstrates the power of the proposed approach in Fig. \ref{fig: conceptualFlow}(b) over the conventional approach in Fig. \ref{fig: conceptualFlow}(a). Additionally, and importantly, RS also comes at a lower complexity since it only relies on linear precoding. RS is therefore a promising, powerful, and robust non-orthogonal transmission technique for real-world applications.

\par \textit{Second}, this paper shows that, in MISO BC with partial CSIT, one can get even better rate region performance than linearly precoded RS (and DPC) by marrying RS and DPC, and using DPC to encode the private parts of the messages. This leads to another transmission strategy, denoted as Dirty Paper Coded Rate-Splitting (DPCRS). We show that the  rate region of DPCRS in MISO BC with partial CSIT is enlarged beyond that of conventional DPC and that of linearly precoded RS. Gaining benefits from the capability of RS to partially decode the interference and partially treat interference as noise, DPCRS is less sensitive to the variation of CSIT inaccuracies, network loads, and user deployments compared with DPC and other existing transmission strategies.

\par \textit{Organization:} The rest of the paper is organized as follows. The system
model is described in Section \ref{sec: system model}. The problem formulation for  the proposed strategies is specified in Section \ref{sec: problem formulation}. 
In Section \ref{sec: algorithm}, the proposed algorithm is described followed by the numerical results in Section \ref{sec: numerical results}. Finally, conclusion is made  in Section \ref{sec: conclusion}.

\par \textit{Notations:} Bold lower and upper case letters denote vectors and matrices, respectively. $\|\cdot\|$ represents Euclidean norm. The notations $(\cdot)^H$, $(\cdot)^T$, $\mathrm{tr}(\cdot)$, $\mathbb{E}\{\cdot\}$ respectively denote the Hermitian, transpose, trace and expectation operators. $\mathbf{I}$ denotes the identity matrix. $\sim$ denotes ``distributed as” and $\mathcal{CN}(0,\sigma^2 )$ denotes the Circularly Symmetric Complex Gaussian (CSCG) distribution with zero mean and variance $\sigma^2$. 
The  notations  in the system model are summarized in Table \ref{tab: notation}.

\begin{table*}[t!]
	\centering
	\caption{Notations}
	\label{tab: notation}	
		\begin{tabular}{|l|l|}
			\hline
			\textbf{Notation}                                      & \textbf{Description }                                                   \\ \hline
		$N_t$                                     & number of transmit antennas                               \\ \hline
		$K$                           &   number of users                               \\ \hline
		$\mathcal{K}$                             &  set of users \\ \hline
		$P_t$                             &  transmit power limit \\ \hline
		$\sigma_{e,k}^2$                             &  variance of error \\ \hline
			$\alpha$                             &  quality of CSIT in the high SNR regime \\ \hline
				$\pi$                             & DPC  encoding order  \\ \hline
			$\pi'$                             &   decoding order of partial-common streams  \\ \hline
	$\mathcal{I}_k$                             &    index set of all the streams  to be decoded at user-$k$  \\ \hline
	$\pi'_k$                             &    index set of partial-common streams to be decoded  at user-$k$ based on decoding order $\pi'$ \\ \hline
	$\mathcal{K}_c$                             &    index set of all common streams  \\ \hline
	$\bar{\mathcal{K}}_{c,k}^{i}$                             &    index set of all undecoded common streams at user-$k$ when decoding the stream $s_i$  \\ \hline
	$\bar{\mathcal{K}}_k$                             &    index set of all undecoded streams at user-$k$ when decoding the private stream $s_k$  \\ \hline
	${\mathcal{K}}_i$                             &    set of users to decode the common stream $s_i$  \\ \hline
	$\overline{R}_{i,k}^{\textrm{x}}$                             &   ER at user-$k$ to decode stream $s_i$ for strategy ``\textrm{x}" \\ \hline
	$\overline{R}_{k,tot}^{\textrm{x}}$                             &   total ER at user-$k$  for strategy ``\textrm{x}" \\ \hline
	$\overline{C}_{k}$                             &   ER of the common stream $s_c$ allocated to user-$k$ for 1-DPCRS  \\ \hline
	$\overline{C}_{k}^i$                             &   ER of the common stream $s_i$ allocated to user-$k$ for M-DPCRS  \\ \hline
	\end{tabular}	
\end{table*}

\section{System Model}
\label{sec: system model}
\par In this work, we consider a  MISO BC with one multi-antenna Base Station (BS) simultaneously serving $K$ single-antenna users. The BS is equipped with $N_t$ transmit antennas and the users are indexed by $\mathcal{K}=\{1,\ldots,K\}$. The signal received by user-$k$  for a given channel use (time or frequency) is 
\begin{equation}
\label{eq: receivedSignalModel}
	y_k=\mathbf{h}_k^H\mathbf{x}+n_k, \forall k\in \mathcal{K}, 
\end{equation}
where $\mathbf{h}_k\in\mathbb{C}^{{N_t}}$ is the channel between the BS and user-$k$. $\mathbf{x}\in\mathbb{C}^{{N_t}}$ is the signal vector transmitted in a given channel use subject to the transmit power constraint $\mathbb{E}\{\|\mathbf{x}\|^2\}\leq P_t$. 
$n_k \sim \mathcal{CN}(0,\sigma_{n,k}^2 )$ is the Additive White Gaussian Noise (AWGN). Without loss of generality, we assume that $\sigma_{n,k}^2=\sigma_{n}^2=1,\forall k\in\mathcal{K}$. Hence, the transmit SNR defined as $\textrm{SNR}\triangleq\frac{P_t}{\sigma_{n}^2 }$ is equal to  $ P_t$.

\subsection{Channel Model}
\label{sec: channelModel}
\par Due to  the uplink channel estimation error caused by  quantized  feedback  \cite{NJindalMIMO2006},  feedback delay  \cite{doppler2010Caire,DoF2013SYang}, etc, CSIT  is commonly imperfect. 
In this work, we assume  perfect CSIR and partial CSIT, which is modeled by
\begin{equation}
	\mathbf{H}=\widehat{\mathbf{H}}+\widetilde{\mathbf{H}},
\end{equation}
where $\mathbf{H}=[\mathbf{h}_1,\dots,\mathbf{h}_K]$ is the actual CSI known at all users. $\widehat{\mathbf{H}}=[\widehat{\mathbf{h}}_1,\dots,\widehat{\mathbf{h}}_K]$ is the estimated CSI known at the BS. $\widetilde{\mathbf{H}}=[\widetilde{\mathbf{h}}_1,\dots,\widetilde{\mathbf{h}}_K]$ is the CSIT estimation error matrix with each element of the $k$th-column for user-$k$ characterized  by an independent and identically distributed (i.i.d.) zero-mean complex Gaussian distribution variable with $\mathbb{E}\{\widetilde{\mathbf{h}}_k \widetilde{\mathbf{h}}_k^H\}=\sigma_{e,k}^2\mathbf{I}$. The variance of the error $\sigma_{e,k}^2$ is assumed to scale exponentially with SNR as $\sigma_{e,k}^2\sim O(P_t^{-\alpha})$, where $\alpha\in[0,\infty)$ is the quality scaling factor  interpreted as the quality of CSIT  in the high SNR regime \cite{AG2015,NJindalMIMO2006,doppler2010Caire,DoF2013SYang,RS2016hamdi}. $\alpha=0$ represents partial CSIT with finite precision, e.g. a constant number of feedback bits, while $\alpha=\infty$ represents  perfect CSIT.
The joint distribution of $\{ \mathbf{H},\widehat{\mathbf{H}}\}$  is assumed to be stationary and ergodic \cite{RS2016hamdi}.   ${\mathbf{H}}$ over the entire transmission is unknown at the BS but the conditional density $f_{{\mathbf{H}}|\widehat{\mathbf{H}}}({\mathbf{H}}|\widehat{\mathbf{H}})$ is assumed to be known at the BS.

\subsection{Conventional Dirty Paper Coding}
\label{sec: DPC}
\par In the conventional DPC \cite{DPC1983,jindal2002duality, duality2003Andrew, DPCrateRegion03Goldsmith}, with a certain encoding order ${\pi}$ (where ${\pi}$ defined as ${\pi}\triangleq[\pi(1),\ldots,\pi(K)]$ is a permutation of \{$1,\ldots,K$\} such that the message $W_{\pi(i)}$ is encoded before $W_{\pi(j)}$ if $i<j$),  the BS  starts encoding from  message $W_{\pi(1)}$  for user-${\pi(1)}$  to  message $W_{\pi(K)}$ for user-${\pi(K)}$. The messages  are encoded into a set of symbol streams $s_{\pi(1)},\ldots,s_{\pi(K)}$  to be transmitted for a given channel use. The stream vector $\mathbf{s}\triangleq[s_{\pi(1)},\ldots,s_{\pi(K)}]^T$ is precoded by $\mathbf{P}\triangleq[\mathbf{p}_{\pi(1)},\ldots,\mathbf{p}_{\pi(K)}]$, where $\mathbf{p}_{\pi(k)}\in\mathbb{C}^{N_t}$ is the precoder for user-${\pi(k)}$, and the resulting superposed transmit signal is 
\begin{equation}
\mathbf{x}=\mathbf{P}\mathbf{s}={{\sum_{k\in\mathcal{K}}\mathbf{p}_{\pi(k)}{s}_{\pi(k)}}}.
\end{equation}
Assuming CSCG inputs with $\mathbb{E}\{\mathbf{s}\mathbf{s}^H\}=\mathbf{I}$, the transmit power constraint is equivalent to $\mathrm{tr}(\mathbf{P}\mathbf{P}^H)=P_t$. If CSIT is perfect,   the encoded data stream $s_{\pi(k)}$ experiences no interference from  previously encoded data streams $\{s_{\pi(i)}|i<k\}$ according to the principle of DPC \cite{DPCrateRegion03Goldsmith}. However, as the BS has no access to the  exact channel ${\mathbf{H}}$, $\mathbf{P}$ is designed at the BS based on the estimated channel state $\widehat{\mathbf{H}}$. Only part of the interference from $\widehat{\mathbf{h}}_{\pi(k)}^H\sum_{i<k}\mathbf{p}_{\pi(i)}s_{\pi(i)}$ is removed from the   signal received at user-$\pi(k)$. The resulting received signal is given by
\begin{equation}
	\small
	y_{\pi(k)}=\widetilde{\mathbf{h}}_{\pi(k)}^H\sum_{i<k}\mathbf{p}_{\pi(i)}s_{\pi(i)}+{\mathbf{h}}_{\pi(k)}^H\sum_{j\geq k}\mathbf{p}_{\pi(j)}s_{\pi(j)}+n_{\pi(k)}.
\end{equation}
Each user directly decodes the intended message by treating any residual interference as noise. As the precoders are designed at the BS based on the  channel estimate $\widehat{\mathbf{H}}$ and each user decodes the intended stream based on the exact channel ${\mathbf{H}}$, the instantaneous  rate  of decoding $s_{\pi(k)}$ at user-$\pi(k)$ is determined by one   joint fading state $\{ \mathbf{H},\widehat{\mathbf{H}}\}$ given as
\begin{equation}
\label{eq: rate DPC}
	\begin{aligned}
&\resizebox{.12\textwidth}{!} {$R_{\pi(k)}^{\textrm{DPC}}( \mathbf{H},\widehat{\mathbf{H}})=$}\\
&\resizebox{.446 \textwidth}{!} {$\log_2\left(1+\frac{|{\mathbf{h}}_{\pi(k)}^H\mathbf{p}_{\pi(k)}|^2}{\sum_{i<k}|\widetilde{\mathbf{h}}_{\pi(k)}^H\mathbf{p}_{\pi(i)}|^2+\sum_{j> k}|\mathbf{h}_{\pi(k)}^H\mathbf{p}_{\pi(j)}|^2+1}\right).$}
	\end{aligned}
	\end{equation}
As the BS only knows the channel estimate $\widehat{\mathbf{H}}$ without any knowledge of  the exact channel ${\mathbf{H}}$, $R_{\pi(k)}^{\textrm{DPC}}( \mathbf{H},\widehat{\mathbf{H}})$ may be overestimated and unachievable at user-$\pi(k)$ \cite{RS2016hamdi}.   A more robust approach is to design the precoders at the BS based on the Ergodic Rate (ER) under the assumption that the transmission is delay-unlimited. 
The ER characterizes the long-term performance of user-$\pi(k)$ over  all possible joint fading states $\{ \mathbf{H},\widehat{\mathbf{H}}\}$, which is defined as
\begin{equation}
\label{eq: ER DPC}
\overline{R}_{\pi(k),tot}^{\textrm{DPC}}\triangleq\mathbb{E}_{\{ \mathbf{H},\widehat{\mathbf{H}}\}}\left\{R_{\pi(k)}^{\textrm{DPC}}( \mathbf{H},\widehat{\mathbf{H}})\right\}.
\end{equation}

\subsection{ Dirty Paper Coded Rate-Splitting}
\label{sec: DPCRS}
\subsubsection{Motivation}
\par
\begin{figure}[t!]
	\centering
	\includegraphics[width=2.8in]{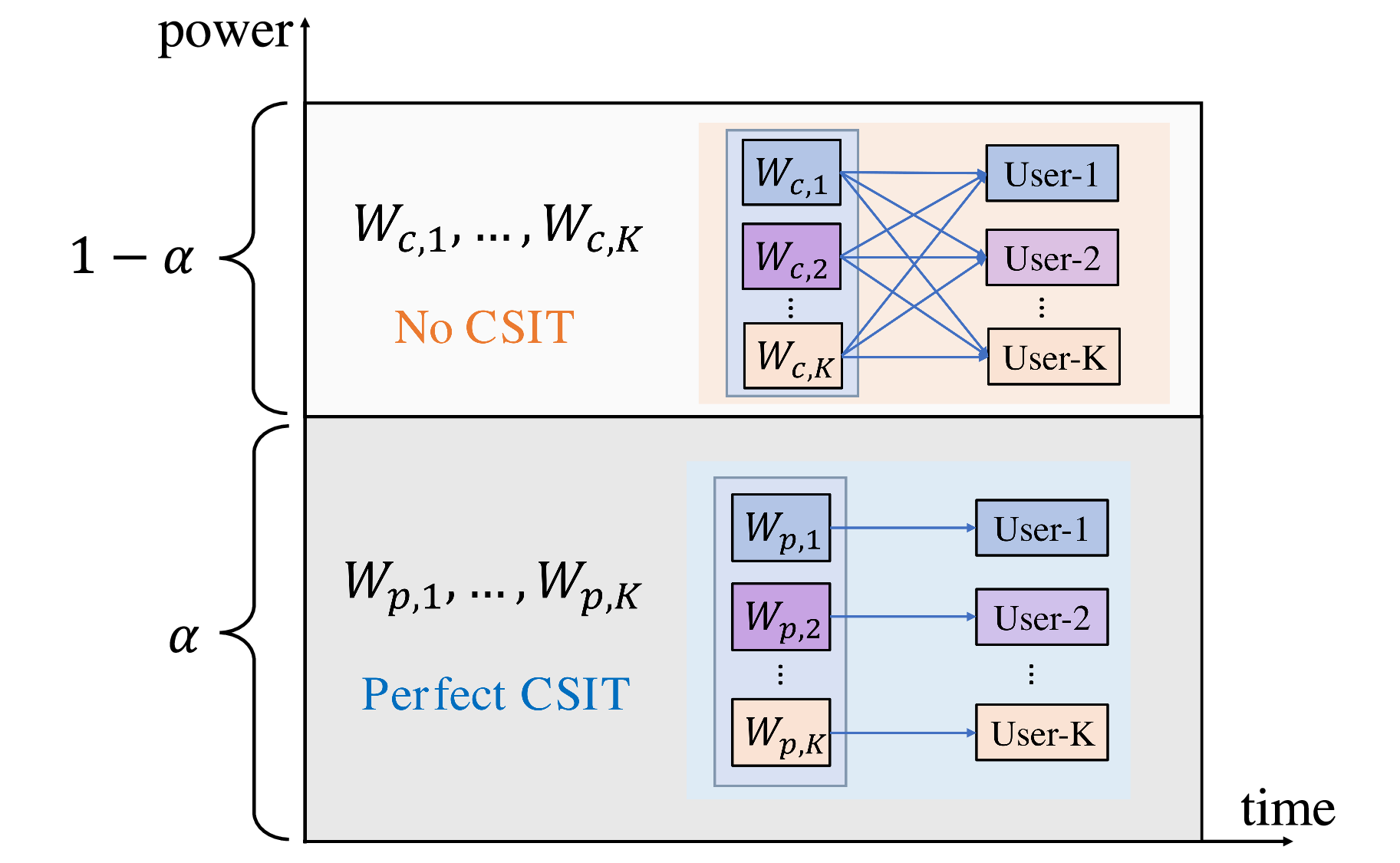}%
	\caption{Weighted sum interpretation of an underloaded $K$-user MISO BC with partial CSIT.}
	\label{fig: motivation}
\end{figure}

 The sum DoF achieved by RS in an underloaded ($N_t\geq K$) $K$-user MISO BC with partial CSIT is given by $1+(K-1)\alpha$ \cite{RS2016hamdi}, where $\alpha$ is the quality scaling factor as defined in Section \ref{sec: channelModel}. This sum DoF matches the upper bound obtained from the Aligned Image Sets in \cite{AG2015}. As a consequence, RS achieves the optimal DoF in this setting. This contrasts with the conventional approach of Fig. \ref{fig: conceptualFlow}(a) that achieves a sum DoF of $\max\{1,K\alpha\}$ \cite{RS2016hamdi}. Interestingly, $1+(K-1)\alpha$ can be equivalently written as $(1-\alpha)+K\alpha$. Leveraging the weighted-sum interpretation in \cite{chenxi2013OFDMA,chenxi2017brunotopology} and the notion of signal-space partitioning in \cite{GDoF2018TIT, Yuan2016MIMOIC}, one can interpret $(1-\alpha)+K\alpha$ as the DoF achieved by the superposition of two sub-networks in the power domain: a first sub-network consisting of a \textit{$K$-user MISO BC with perfect CSIT} using a power level $\alpha$ contributing to a sum DoF of $K\alpha$, and a second sub-network consisting of a \textit{$K$-user MISO BC with no CSIT} using the remaining power level $1-\alpha$ contributing to a sum DoF of $1-\alpha$, as illustrated in Fig. \ref{fig: motivation}. Loading data onto those two sub-networks is achieved by an non-orthogonal transmission in the power domain using RS that splits messages into common and private parts, with the private parts loaded onto the first sub-network and the common parts onto the second sub-network. Since the first sub-network can be viewed as a $K$-user MISO BC with perfect CSIT, and DPC is capacity-achieving for such a scenario, it motivates us to encode the private parts using DPC. This leads to the  Dirty Paper Coded RS discussed in the sequel. 

\subsubsection{One-Layer  Dirty Paper Coded Rate-Splitting (1-DPCRS)} 
\label{sec: 1layerDPCRS}
\par
\begin{figure}
	\centering
	\begin{subfigure}[b]{0.46\textwidth}
		\centering
		\includegraphics[width=\textwidth]{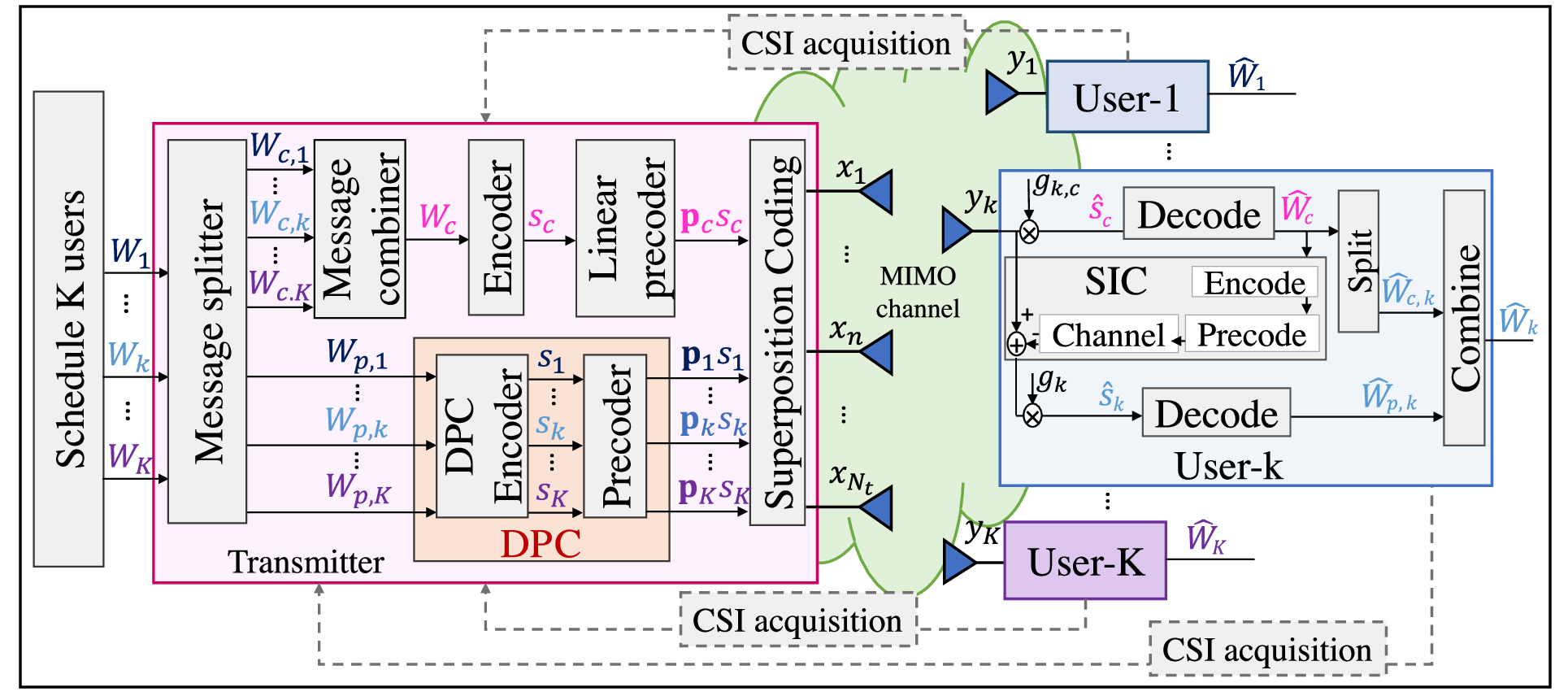}%
		\caption{ $K$-user 1-DPCRS.}	
	\end{subfigure}%
	~\\
	\centering
	\begin{subfigure}[b]{0.48\textwidth}
		\centering
		\includegraphics[width=\textwidth]{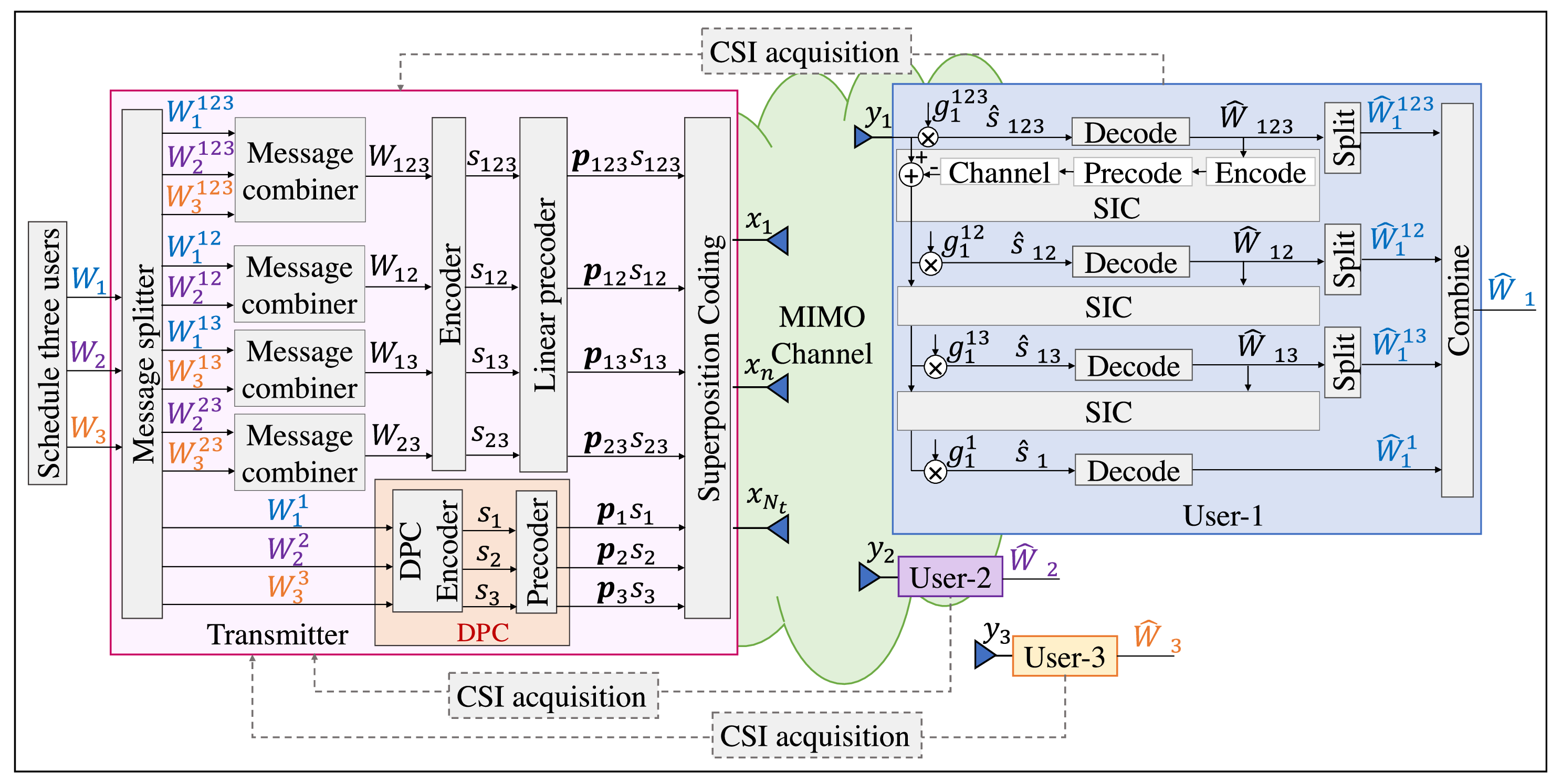}%
		\caption{$3$-user M-DPCRS, $\pi'=12\rightarrow13\rightarrow23$.}
	\end{subfigure}
	\caption{System architecture of  Dirty Paper Coded RS for MISO BC.}
	\label{fig: DPCRS}
\end{figure}

 Though DPC achieves the capacity region of MISO BC with perfect CSIT \cite{jindal2002duality, duality2003Andrew, DPCrateRegion03Goldsmith},  it is sensitive to the CSIT inaccuracy {\cite{DPC2012NanLee, yang2005impact}}.   Motivated by the DoF interpretation at infinite SNR in Fig. \ref{fig: motivation}  and the recent benefits of RS  in multi-antenna BC, in this work, we focus on finite SNR regime  where we first marry 1-layer RS with DPC so as to combat  performance losses of DPC resulting from partial CSIT and  explore a larger   rate region in MISO BC with partial CSIT. The proposed ``\textit{1-layer Dirty Paper Coded RS  (1-DPCRS)}" strategy is illustrated in Fig. \ref{fig: DPCRS}(a).

\par In 1-DPCRS,  message  $W_k$ intended for user-$k, \forall k\in\mathcal{K}$ is first split into one common part $W_{c,k}$ and one private part $W_{p,k}$. The common parts $W_{c,1},\ldots,W_{c,K}$ of all users are combined into the common message $W_c$ and encoded into the common stream $s_c$ to be decoded by all  users for a given channel use. Different from the linearly precoded 1-layer RS strategy studied in the literature  
\cite{RS2016hamdi,RSintro16bruno,mao2017rate,mao2018rate}, the private parts $W_{p,1},\ldots,W_{p,K}$  are  encoded and precoded by DPC  with a certain encoding order $\pi$ into the private streams $s_{\pi(1)},\ldots, s_{\pi(K)}$ to be decoded by the  corresponding users only. The data vector $\mathbf{s}\triangleq[s_c,s_{\pi(1)},\ldots,s_{\pi(K)}]^T$ is precoded by  $\mathbf{P}\triangleq[\mathbf{p}_c,\mathbf{p}_1,\ldots,\mathbf{p}_K]$,  the resulting  transmit signal is 
\begin{equation}
\label{eq: transmit signal DPCRS}
\mathbf{x}=\mathbf{P}\mathbf{s}={{\mathbf{p}_c{s}_c}}+{{\sum_{k\in\mathcal{K}}\mathbf{p}_{\pi(k)}{s}_{\pi(k)}}}.
\end{equation}
The transmit power constraint is  $\mathrm{tr}(\mathbf{P}\mathbf{P}^H)=P_t$ and  CSCG inputs with $\mathbb{E}\{\mathbf{s}\mathbf{s}^H\}=\mathbf{I}$ are assumed.

\par At user sides,  user-$\pi(k)$ first decodes the common stream $s_c$ into $\widehat{W}_c$ by treating  the  interference from all  private streams as noise. With the assist of SIC\footnote{In this work, we only consider the SIC receiver architecture as widely used in the existing works \cite{RS2016hamdi,enrico2017bruno,hamdi2019spawc,chenxi2017brunotopology,chenxi2017bruno,mao2017rate,hamdi2016robust,mao2018EE,Medra2018SPAWC,Lu2018MMSERS,Flores2018ISWCS,hamdi2017bruno,Minbo2016MassiveMIMO,minbo2017mmWave,AP2017bruno,mao2018networkmimo,alaa2019IEEEaccess,mao2019swipt,mao2019TCOM,jian2019CRS,mao2019maxmin}. Other forms of receiver  are worth to be investigated  in  future works to further enhance the rate performance of systems, i.e.,  beamforming with JD  studied in \cite{DPC2012NanLee} and RS with JD studied in \cite{sheng2020RSJD}. The receiver architecture of the proposed DPCRS strategy in this work can also be enhanced by using the JD receiver architecture.}, the decoded common message $\widehat{W}_c$ then goes through the process of re-encoding, precoding, and subtracting from the received signal. 
After decoding the common stream, user-$\pi(k)$ then decodes the intended private stream $s_{\pi(k)}$ into $\widehat{W}_{p,\pi(k)}$ by treating the interference from the  private streams encoded after $s_{\pi(k)}$  as noise. Once $\widehat{W}_c$ and $\widehat{W}_{p,\pi(k)}$ are decoded, user-$\pi(k)$ reconstructs the original message by extracting $\widehat{W}_{c,\pi(k)}$ from $\widehat{W}_c$,  and then combines $\widehat{W}_{c,\pi(k)}$ with $\widehat{W}_{p,\pi(k)}$  into  $\widehat{W}_{\pi(k)}$. The instantaneous rates of decoding the common stream $s_c$  and the private stream $s_{\pi(k)}$ at user-${\pi(k)}$  are  
\begin{subequations}
\begin{align}
&\resizebox{.17 \textwidth}{!} {$R_{c,\pi(k)}^{\textrm{1-DPCRS}}( \mathbf{H},\widehat{\mathbf{H}})=\log_2$}\resizebox{.26\textwidth}{!} {$\left(1+\frac{|{\mathbf{h}}_{\pi(k)}^H\mathbf{p}_{c}|^2}{\sum_{j\in\mathcal{K}}|\mathbf{h}_{\pi(k)}^H\mathbf{p}_{\pi(j)}|^2+1}\right)$}, \label{eq: DPC common 1-RS}\\
&\resizebox{.14 \textwidth}{!} {$R_{\pi(k)}^{\textrm{1-DPCRS}}( \mathbf{H},\widehat{\mathbf{H}})=$} \nonumber\\
&\resizebox{.03 \textwidth}{!} {$\log_2$}\resizebox{.43 \textwidth}{!} {$ \left(1+\frac{|{\mathbf{h}}_{\pi(k)}^H\mathbf{p}_{\pi(k)}|^2}{\sum_{i<k}|\widetilde{\mathbf{h}}_{\pi(k)}^H\mathbf{p}_{\pi(i)}|^2+\sum_{j> k}|\mathbf{h}_{\pi(k)}^H\mathbf{p}_{\pi(j)}|^2+1}\right).$} \label{eq: DPC private 1-RS}
\end{align}
\end{subequations}
The respective ERs of decoding $s_c$ and $s_{\pi(k)}$ at user-${\pi(k)}$ using 1-DPCRS are defined as 
\begin{equation}
\begin{aligned}
&\overline{R}_{c,\pi(k)}^{\textrm{1-DPCRS}}\triangleq\mathbb{E}_{\{\mathbf{H},\widehat{\mathbf{H}}\}}\{R_{c,\pi(k)}^{\textrm{1-DPCRS}}( \mathbf{H},\widehat{\mathbf{H}})\},\\
&\overline{R}_{\pi(k)}^{\textrm{1-DPCRS}}\triangleq\mathbb{E}_{\{ \mathbf{H},\widehat{\mathbf{H}}\}}\{R_{\pi(k)}^{\textrm{1-DPCRS}}( \mathbf{H},\widehat{\mathbf{H}})\}.
\end{aligned}
\end{equation}
To ensure $s_c$ is successfully decoded by all users, the ER of the common stream $s_c$ should not exceed
\begin{equation}
\label{eq: 1-DPCRS common}
\overline{R}_{c}^{\textrm{1-DPCRS}}\triangleq\min\left\{\overline{R}_{c,\pi(k)}^{\textrm{1-DPCRS}}\mid k\in \mathcal{K}\right\}. 
\end{equation}
As the common stream is shared by all users, by denoting the ER of  the common stream allocated to user-$\pi(k)$ as $\overline{C}_{\pi(k)}$, we have $\sum_{k\in\mathcal{K}}\overline{C}_{\pi(k)}=\overline{R}_{c}^{\textrm{1-DPCRS}}$.
Therefore,  the total ER of each user using  1-DPCRS is  $\overline{R}_{\pi(k),tot}^{\textrm{1-DPCRS}}=\overline{C}_{\pi(k)}+\overline{R}_{\pi(k)}^{\textrm{1-DPCRS}}$.

\subsubsection{Multi-Layer Dirty Paper Coded Rate-Splitting (M-DPCRS)}
\par To further exploit a larger  rate region for MISO BC with partial CSIT, we incorporate the generalized RSMA framework proposed in \cite{mao2017rate} with DPC and propose a novel transmission scheme, namely, ``\textit{Multi-layer Dirty Paper Coded RS (M-DPCRS)}".  We claim that the proposed  M-DPCRS has a larger  rate region  than DPC and other linearly precoded schemes  in  MISO BC with partial CSIT.

\par Compared with the linearly precoded RSMA framework proposed in \cite{mao2017rate}, the main difference of the proposed strategy comes from the non-linear DPC encoding and precoding for the private streams. How the user messages are split and combined follows the framework in \cite{mao2017rate}. To simplify the explanation, we introduce a three-user M-DPCRS, where the users are indexed as $\mathcal{K}=\{1,2,3\}$. It can be easily generalized to the $K$-user case if readers understand the three-user M-DPCRS as well as the two-user 1-DPCRS.  As illustrated in Fig. \ref{fig: DPCRS}(b),  user message $W_k$ of user-$k$ is split into four parts at the BS as $\{W_k^{i}|i\in\mathcal{I}_k\}$, where $\mathcal{I}_1=\{123,12,13,1\}$, $\mathcal{I}_2=\{123,12,23,2\}$, $\mathcal{I}_3=\{123,13,23,3\}$. This is different from 1-DPCRS described in Section \ref{sec: 1layerDPCRS} where the message of each user is only split into two parts as $\{W_{c,k},W_{p,k}\}$. The intention of splitting user messages into more different parts is to  form more layers of common streams for different users, so as to manage interference and disparity of channel strengths more flexibly. The sub-messages $\{W_k^{123}|k\in\mathcal{K}\}$ are jointly encoded into the common stream $s_{123}$, which is decoded by all the three users. The sub-messages $\{W_1^{12},W_2^{12}\}$ and $\{W_1^{13},W_3^{13}\}$ and $\{W_2^{23},W_3^{23}\}$ are respectively encoded into the partial-common streams $s_{12}$, $s_{13}$, $s_{23}$ to be decoded by the intended two users. The private messages $W_k^k,\forall k\in\mathcal{K}$ are  encoded via DPC with encoding order $\pi$ into the private streams $s_{\pi(1)}, s_{\pi(2)}, s_{\pi(3)}$ for the respective user only. We assume the private stream of user-$\pi(k)$ is encoded after user-$\pi(i)$ if $k>i$. Define the common stream index set as  $\mathcal{K}_c\triangleq\{123,12,13,23\}$. The encoded stream vector  $\mathbf{s}\triangleq\{s_{k}\mid k\in\mathcal{K}_c\cup\mathcal{K}\}$ is  precoded by $\mathbf{P}\triangleq\{\mathbf{p}_{k}\mid k\in\mathcal{K}_c\cup\mathcal{K}\}$. The resulting transmit signal $\mathbf{x}\in\mathbb{C}^{N_t}$ is
\begin{equation}
\label{eq: MDPCRS transmitSig}
\mathbf{x}=\mathbf{P}\mathbf{s}={{\sum_{i\in\mathcal{K}_c}\mathbf{p}_is_i}}+{{\sum_{k\in\mathcal{K}}\mathbf{p}_{\pi(k)}\mathbf{s}_{\pi(k)}}}.
\end{equation} 

\par In the three-user M-DPCRS strategy, each user requires three layers of SIC to sequentially decode and remove the intended common streams before decoding the intended private stream. The decoding order of the common streams  follows the rule  that the data streams intended for a larger number of users have higher decoding priorities.  Such rule is commonly adopted in the literature of RS \cite{mao2017rate,Minbo2016MassiveMIMO,mao2019TCOM}, and is motivated by the DoF results and analysis \cite{RS2016hamdi,enrico2017bruno}. Therefore,  $s_{123}$ is decoded first at all users, followed by the intended partial-common streams $s_{12}, s_{13}, s_{23}$. The private streams $s_{1}, s_{2}, s_{3}$ are decoded at the end. As each of the partial-common streams $s_{12}, s_{13}, s_{23}$ are to be decoded by two users, their decoding order is considered to be jointly optimized with the precoders in this work. We define $\pi'\triangleq[\pi'(1), \pi'(2), \pi'(3)]$ as one permutation  of  $\{12,13,23\}$ such that $s_{\pi'(i)}$ is decoded before $s_{\pi'(j)}$ if $i<j$ at all users\footnote{Notice that $\pi'$ is different from $\pi$. $\pi'$ is the decoding order of the linearly precoded common streams at all users while $\pi$ is the encoding order of the DPC-coded private streams at the BS.}. As each user only decodes two of the partial-common streams, we further define $\pi'_k\triangleq[\pi_k'(1), \pi_k'(2)]$ as the two partial-common streams to be decoded at user-$k$ based on the decoding order $\pi'$. For example, when $\pi'=[12,13,23]$ as illustrated in Fig. \ref{fig: DPCRS}(b), the decoding order at user-1 is $\pi'_{1}=[12,13]$.  $s_{\pi'_{1}(1)}= s_{12}$ is decoded first followed by $s_{\pi'_{1}(2)}= s_{13}$. Hence, user-$\pi(k)$ sequentially decodes  the  streams $\{s_{i}|i\in\mathcal{I}_{\pi(k)}\}$ for a given DPC encoding order $\pi$ of the private streams and a given decoding order  $\pi'$ of the partial-common streams.  The instantaneous rates of  decoding  the common streams $s_{i}, i\in \{123, \pi'_{\pi(k)}\}$ and the private stream $s_{\pi(k)}$ at user-$\pi(k)$ are
\begin{subequations}
\begin{align}
&\resizebox{.14 \textwidth}{!} {$R_{i,\pi(k)}^{\textrm{M-DPCRS}}( \mathbf{H},\widehat{\mathbf{H}})=$}\\
&\resizebox{.03 \textwidth}{!} {$\log_2$}\resizebox{.29 \textwidth}{!} {$\left(1+\frac{|{\mathbf{h}}_{\pi(k)}^H\mathbf{p}_{i}|^2}{\sum_{j\in\bar{\mathcal{K}}_{c,\pi(k)}^{i}\cup\mathcal{K}}|\mathbf{h}_{\pi(k)}^H\mathbf{p}_{j}|^2+1}\right)$},\forall i\in \{123, \pi'_{\pi(k)}\},  \label{eq: DPC common M-RS}\\
&\resizebox{.14 \textwidth}{!} {$R_{\pi(k)}^{\textrm{M-DPCRS}}( \mathbf{H},\widehat{\mathbf{H}})=$} \nonumber\\
&\resizebox{.03 \textwidth}{!} {$\log_2$}\resizebox{.39 \textwidth}{!} {$\left(1+\frac{|{\mathbf{h}}_{\pi(k)}^H\mathbf{p}_{\pi(k)}|^2}{\sum_{j\in\bar{\mathcal{K}}_{\pi(k)}}|\mathbf{h}_{\pi(k)}^H\mathbf{p}_{j}|^2+\sum_{j<k}|\widetilde{\mathbf{h}}_{\pi(k)}^H\mathbf{p}_{\pi(j)}|^2+1}\right)$}, \label{eq: DPC private M-RS}
\end{align}
\end{subequations}
where $\bar{\mathcal{K}}_{c,\pi(k)}^i$ is the set of remaining undecoded common streams when  user-$\pi(k)$ decodes the common stream $s_i$, i.e., $\bar{\mathcal{K}}_{c,\pi(k)}^{123}=\mathcal{K}_c\setminus\{123\}$.
 $\bar{\mathcal{K}}_{\pi(k)}=\mathcal{K}_c\setminus\{123, \pi'_{\pi(k)}\}\cup\{\pi(j)|j>k\}$ is  the set of undecoded  common and private streams at user-$\pi(k)$ when decoding the intended private stream $s_{\pi(k)}$.

\par In the three-user M-DPCRS,  the ERs of decoding the intended the common and private streams $\{s_i\mid i\in \mathcal{I}_{\pi(k)}\}$ at user-$\pi(k)$ are defined as 
\begin{equation}
\overline{R}_{i,\pi(k)}^{\textrm{M-DPCRS}}\triangleq\mathbb{E}_{\{ \mathbf{H},\widehat{\mathbf{H}}\}}\{R_{i,\pi(k)}^{\textrm{M-DPCRS}}( \mathbf{H},\widehat{\mathbf{H}})\},
\end{equation} where $R_{i,\pi(k)}^{\textrm{M-DPCRS}}( \mathbf{H},\widehat{\mathbf{H}})=R_{\pi(k)}^{\textrm{M-DPCRS}}( \mathbf{H},\widehat{\mathbf{H}})$ and $\overline{R}_{i,\pi(k)}^{\textrm{M-DPCRS}}$ is simplified to $\overline{R}_{\pi(k)}^{\textrm{M-DPCRS}}$ when $i=\pi(k)$. To ensure the common streams $\{s_{i}\mid i\in\mathcal{K}_c\}$ are successfully decoded by the intended users, the  ERs of the common streams should not exceed $\overline{R}_{i}^{\textrm{M-DPCRS}}\triangleq\min\{\overline{R}_{i,\pi(k)}^{\textrm{M-DPCRS}}\mid \pi(k)\in \mathcal{K}_i\}$, where $\mathcal{K}_i$  denotes the group of users decoding the common stream $s_i$.   For instance, $\mathcal{K}_i=\{1,2\}$ when $i=12$ and $\mathcal{K}_i=\mathcal{K}$ when $i=123$. By introducing scalar variable $\overline{C}_{k}^{i}$ to denote the ER of common stream $s_i$ allocated to user-$k$,
we have
$\sum_{k\in\mathcal{K}_i}\overline{C}_{k}^{i}=\overline{R}_{i}^{\textrm{M-DPCRS}}$.  As the message of each user is split into four parts,  the ER of user-$\pi(k)$ is the sum of the rate allocated to user-$\pi(k)$ in the intended common streams, i.e.,
$\overline{R}_{ \pi(k),tot}^{\textrm{M-DPCRS}}=\sum_{i\in\mathcal{I}_{\pi(k)}\setminus\{\pi(k)\}}\overline{C}_{\pi(k)}^{i}+\overline{R}_{\pi(k)}^{\textrm{M-DPCRS}}.$

\begin{remark}
\par As the common streams of RS are linearly precoded, the achievability of ERs for the common streams based on (\ref{eq: DPC common 1-RS}) and (\ref{eq: DPC common M-RS}) is guaranteed. In contrast, the achievability of the ERs for the private streams calculated by (\ref{eq: DPC private 1-RS}), (\ref{eq: DPC private M-RS}) and (\ref{eq: rate DPC}) is  unknown since the private streams are encoded based on DPC and there is no prior work  that rigorously show the  achievability of the ERs of DPC for multi-antenna BC with partial CSIT to the best of our knowledge. 
If the ER based on (\ref{eq: rate DPC}) is achievable,   ERs based on (\ref{eq: DPC private 1-RS}) and (\ref{eq: DPC private M-RS}) are achievable, and vice versa. This is due to the fact that DPC is only applied to the encoding and precoding of the private streams in both 1-DPCRS and M-DPCRS. In the following, we explain the achievability issue in detail and discuss its potential influence to the study of DPCRS in this work. 

\par DPC  based on the Gel'fand-Pinsker coding applied on the additive channel is given as  $Y=GX+S+Z$, where $X$ is the channel input, $G$ is the effective channel gain, $S$ is the interference, and $Z$ is AWGN \cite{DPC1983}, \cite[Section 1.4.2]{DPC2011book}. When CSIT is perfect, $G$  and $S$ are known at the transmitter and $Z$ is independent on $S$.   $S$ is  successfully pre-cancelled at the transmitter. User rates of DPC with perfect CSIT are therefore achievable. However, when CSIT is imperfect, $G$ is not known at the transmitter and $Z$ is dependent on $S$. Whether $S$ can be  pre-cancelled is unknown and  the  achievability issue arises.

\par In equation (\ref{eq: rate DPC}), $G,  X,  S, Z$ are respectively equivalent to \resizebox{.13 \textwidth}{!} {${\mathbf{h}}_{\pi(k)}^H\mathbf{p}_{\pi(k)}\rightarrow G$}, \resizebox{.09 \textwidth}{!}{$s_{\pi(k)}\rightarrow X$}, \resizebox{.21 \textwidth}{!} {$ \widehat{\mathbf{h}}_{\pi(k)}^H\sum_{i<k}\mathbf{p}_{\pi(i)}s_{\pi(i)}\rightarrow S$}, \resizebox{.37\textwidth}{!} {$ \sum_{i<k}|\widetilde{\mathbf{h}}_{\pi(k)}^H\mathbf{p}_{\pi(i)}|^2+\sum_{j> k}|\mathbf{h}_{\pi(k)}^H\mathbf{p}_{\pi(j)}|^2+1\rightarrow Z$}  and we assume that $S$ can be  pre-cancelled at the transmitter. One standpoint for the ER based on (\ref{eq: rate DPC}) to be achievable is that the dependency between $S$ and $Z$ should increase the capacity   according to footnote 2 of \cite{yang2005impact}. By treating $S$ and $Z$ as independent as if $Z$ is  AWGN, we should be getting an achievable performance which ignores (or does not exploit) the dependency. The channel gain $G$ can be obtained at the transmitter by assuming that once the precoders are selected and synchronized at the transmitter and receivers, each receiver feeds back the composite channel $\mathbf{h}_k^H\mathbf{p}_k$, which provides $G$. 

\par Showing  the  achievability of the ERs of DPC for multi-antenna BC with partial CSIT rigorously is beyond the  scope of this work and would deserve a thorough study afterwards.
To complete our discussion, we summarize  the  possible cases of the achievability of ERs based on (\ref{eq: rate DPC}), (\ref{eq: DPC private 1-RS}) and (\ref{eq: DPC private M-RS}), and discuss the merits of DPCRS  to be explored in this work for each case:
	\begin{enumerate}
		\item The first case is when the ER of DPC based on (\ref{eq: rate DPC}) and the ERs of the private streams for DPCRS based on (\ref{eq: DPC private 1-RS}) and (\ref{eq: DPC private M-RS}) are achievable, then we are able to draw the conclusion that the proposed DPCRS boosts the achievable rate region and becomes  a new benchmark for assessing the achievable rate region in multi-antenna BC with partial CSIT.
		\item The second case is when ERs based on (\ref{eq: rate DPC}), (\ref{eq: DPC private 1-RS}) and (\ref{eq: DPC private M-RS})  are not achievable, which are then the respective outer bounds of the achievable rate regions of DPC and DPCRS in MISO BC with partial CSIT. In such case, the performance benefits of linearly precoded RS are further exploited since we show throughout the paper that the achievable rate region of linearly precoded RS outperforms  the outer bound of DPC rate region and is almost overlapped with the outer bound of DPCRS rate region in MISO BC with partial CSIT. 
	\end{enumerate}
	
The rate achievability issue discussed above was the motivation in \cite{yang2005impact} to derive an achievable lower bound on the rate for DPC with imperfect CSIT. To further show the superiority of RS and DPCRS over DPC in imperfect CSIT, following \cite{yang2005impact}, we obtain   lower bounds of (\ref{eq: rate DPC}) and (\ref{eq: DPC private 1-RS})  in the Appendix, and also confirm the superiority of RS and DPCRS over DPC using achievable lower bounds on the rate.
\end{remark}

\section{Problem Formulation}
\label{sec: problem formulation}
\par An intuitive method of precoder design at the BS is to optimize the instantaneous precoder $\mathbf{P}$ based on the knowledge of the estimated channel state $\widehat{\mathbf{H}}$ by maximizing the instantaneous Weighted Sum Rate (WSR) subject to instantaneous power constraint $\mathrm{tr}(\mathbf{P}\mathbf{P}^H)\leq P_t$.  However, the partial CSIT may lead to an undecodable rate \cite{RS2016hamdi}. In this work, we aim at optimizing precoders so as to maximize the Weighted Ergodic Sum Rate (WESR) which captures the long-term (ergodic) weighted sum rate performance of all users. WESR for a given  strategy  $\textrm{x}\in$\{``{DPC}",``{1-DPCRS}",``{M-DPCRS}"\} is defined as 
\begin{equation}
 \textrm{WESR}^{\textrm{x}}\triangleq \sum_{k\in\mathcal{K}}u_{k}  \overline{R}_{k,tot}^{\textrm{x}}.
\end{equation}
 In this section, the precoder and message-split design problems for WESR maximization in  MISO BC with partial CSIT are formulated. Before specifying the formulated problems, we start by answering the question: \textit{how to maximize WESR at the BS with instantaneous imperfect CSIT?}

\par Though  BS is not able to estimate the instantaneous rates, the Average Rates (ARs) of users are predictable at the BS \cite{RS2016hamdi}. The ARs are defined in Definition \ref{def: AR}.
\begin{definition}
	\label{def: AR}
	The AR of decoding the stream $s_i$  at user-$k,k\in\mathcal{K}$ for a given channel estimate $\widehat{\mathbf{H}}$ and precoder $\mathbf{P}(\widehat{\mathbf{H}})$ is given by
	\begin{equation}
	\label{eq: average rate}
	\widehat{R}_{i,k}^{\textrm{x}}(\widehat{\mathbf{H}})\triangleq\mathbb{E}_{\{\mathbf{H}\mid \widehat{\mathbf{H}}\}}\left\{R_{i,k}^{\textrm{x}}(\mathbf{H}, \widehat{\mathbf{H}})\mid \widehat{\mathbf{H}}\right\},
	\end{equation}
	where $R_{i,k}^{\textrm{x}}(\mathbf{H},\widehat{\mathbf{H}})=R_{k}^{\textrm{x}}(\mathbf{H},\widehat{\mathbf{H}})$  and $\widehat{R}_{i,k}^{\textrm{x}}(\widehat{\mathbf{H}})$ is simplified to $\widehat{R}_{k}^{\textrm{x}}(\widehat{\mathbf{H}})$ when $i=k$. $\textrm{x}\in$\{``{DPC}",``{1-DPCRS}",``{M-DPCRS}"\}. 
\end{definition}
\par Notice that AR is a short-term measure for an instantaneous channel estimate $\widehat{\mathbf{H}}$ at the BS. It captures the expected rate over the  CSIT uncertainty for a given  $\widehat{\mathbf{H}}$ and conditional density $f_{{\mathbf{H}}\mid \widehat{\mathbf{H}}}({\mathbf{H}}\mid \widehat{\mathbf{H}})$. This is different from ER  that captures the long-term performance over a long sequence of channel uses $\{\mathbf{H},\widehat{\mathbf{H}}\}$ spanning almost all possible joint channel states. We could then obtain the relation between ER and AR \cite{RS2016hamdi} as provided in Proposition 	\ref{prop: ER vs AR}:
\begin{proposition}
	\label{prop: ER vs AR}
	The ER and AR of decoding the stream $s_i$  at user-$k,k\in\mathcal{K}$  for a given  strategy $\textrm{x}\in$\{``{DPC}",``{1-DPCRS}",``{M-DPCRS}"\} follows the relation:
	\begin{equation}
	\label{eq: AR-ER}
	\overline{R}_{i,k}^{\textrm{x}}=\mathbb{E}_{ \widehat{\mathbf{H}}}\left\{	\widehat{R}_{i,k}^{\textrm{x}}(\widehat{\mathbf{H}})\right\}.
	\end{equation}
\end{proposition}

\par \textit{Proof}: 
Relation (\ref{eq: AR-ER}) is obtained by the law of total expectation and definition of AR in (\ref{eq: average rate}), \resizebox{.18\textwidth}{!} {$\mathbb{E}_{\{\mathbf{H}, \widehat{\mathbf{H}}\}}\left\{R_{i,k}^{\textrm{x}}(\mathbf{H}, \widehat{\mathbf{H}})\right\}=$} \resizebox{.36\textwidth}{!} {$\mathbb{E}_{\widehat{\mathbf{H}}}\left\{\mathbb{E}_{\{\mathbf{H}\mid \widehat{\mathbf{H}}\}}\{R_{i,k}^{\textrm{x}}(\mathbf{H}, \widehat{\mathbf{H}})\mid \widehat{\mathbf{H}}\}\right\}=\mathbb{E}_{ \widehat{\mathbf{H}}}\left\{	\widehat{R}_{i,k}^{\textrm{x}}(\widehat{\mathbf{H}})\right\}$}.\qed

\par Based on Proposition 	\ref{prop: ER vs AR},  the problem of designing the precoder $\mathbf{{P}}$  to maximize the WESR  subject to the short-term rate and power constraints\footnote{Note that the  long-term power constraint, i.e., $\mathbb{E}_{\{\mathbf{H},\widehat{\mathbf{H}}\}}\{\text{tr}(\mathbf{P}\mathbf{P}^{H})\}\leq P_t$ results in  allocating power  across different channel states, which is intractable. Therefore, it is common to replace the long-term
power constraint  with short-term power constraints for each channel state.} is equivalently transformed to the problem of  maximizing Weighted Average Sum Rate (WASR)  defined by $\sum_{k\in\mathcal{K}}u_{\pi(k)}\widehat{R}_{\pi(k),tot}^{\textrm{x}}(\widehat{\mathbf{H}})$  for each  $\widehat{\mathbf{H}}$ due to the fact that  there is no dependencies  in the  objective function and the constraints of the WESR optimization problem among the channel instances. Hence, the  WESR maximization problem is decomposed into the  WASR subproblem for each  $\widehat{\mathbf{H}}$ \cite{RS2016hamdi}. In the following, we specify the formulated WASR maximization problem for all  the considered strategies.

\subsection{DPC}
\label{sec: prob DPC}
\par For a given weight vector $\mathbf{u}=[u_1,\ldots,u_K]$ and a fixed DPC encoding order $\pi$, the   WASR problem for each  $\widehat{\mathbf{H}}$ is:
\begin{subequations}
	\label{eq: DPC b}
	\begin{align}
	&	\max_{\mathbf{{P}}}\,\, \sum_{k\in\mathcal{K}}u_{\pi(k)}\widehat{R}_{\pi(k)}^{\textrm{DPC}}(\widehat{\mathbf{H}})\\
	\mbox{s.t.}\,\,
	&  \,\,\widehat{R}_{\pi(k)}^{\textrm{DPC}}(\widehat{\mathbf{H}})\geq R_{\pi(k)}^{th}, \forall k\in\mathcal{K} \label{c1_DPC b}\\
	&\,\,	\text{tr}(\mathbf{P}\mathbf{P}^{H})\leq P_{t}, \label{c2_DPC b}	
	\end{align}
\end{subequations}
where $\widehat{R}_{\pi(k)}^{\textrm{DPC}}(\widehat{\mathbf{H}})$ is defined in Definition \ref{def: AR} and $R_{{\pi(k)}}^{th}$ is the QoS rate constraint of user-$\pi(k)$. An extra optimization over DPC encoding order $\pi$ is required in order to maximize WESR for a given $\mathbf{u}$. 

\begin{remark}
When CSIT is perfect, the capacity region of MISO BC is achieved by solving problem (\ref{eq: DPC b}) with $R_{\pi(k)}^{th}=0$  for all possible $\pi$ and a set of $\mathbf{u}$. Though it is typical to optimize covariance matrix $\mathbf{Q}_k=\mathbf{p}_k\mathbf{p}_k^H$ in the literature of DPC \cite{DPCregion,DPCrateRegion03Goldsmith}, the optimal covariance matrix of DPC when each user has a single antenna   is rank one  due to  the uplink-downlink  duality \cite{jindal2002duality, duality2003Andrew}. The dual MISO MAC channel has a single transmit antenna at each user and thus a rank one covariance matrix. The transformation to the corresponding covariance matrix in  MISO BC preserves this rank. Hence, the capacity region can be achieved by optimizing the precoder.
\end{remark}

\subsection{1-DPCRS }
\label{sec: prob 1-DPCRS} 
\par Note that the overall ER  of the common stream $\overline{R}_{c}^{\textrm{1-DPCRS}}$ specified in (\ref{eq: 1-DPCRS common}) with minimization outside  the ER of each user  is intractable since the precoders of all fading states are required to be jointly designed. To remove dependency among the channel instances, we consider its lower bound by moving minimization inside of the ER based on the inequality \cite{RS2016hamdi}:
\begin{equation}
\begin{aligned}
\overline{R}_{c}^{\textrm{1-DPCRS}}&=\min_{k\in \mathcal{K}}\left\{\mathbb{E}_{ \widehat{\mathbf{H}}}\left\{	\widehat{R}_{c,\pi(k)}^{\textrm{1-DPCRS}}(\widehat{\mathbf{H}})\right\}\right\} \\
&\geq \mathbb{E}_{ \widehat{\mathbf{H}}}\left\{	\min\left\{ \widehat{R}_{c,\pi(k)}^{\textrm{1-DPCRS}}(\widehat{\mathbf{H}})\mid k\in \mathcal{K}\right\}\right\}.
\end{aligned}
\end{equation}
By defining $\widehat{R}_{c}^{\textrm{1-DPCRS}}(\widehat{\mathbf{H}})\triangleq\min\{\widehat{R}_{c,\pi(k)}^{\textrm{1-DPCRS}}(\widehat{\mathbf{H}})\mid k\in\mathcal{K} \}$ as  the overall AR of the common stream $s_c$  and $\widehat{C}_{\pi(k)}$ as the  AR of the common stream allocated to user-$\pi(k)$, we have $\sum_{k\in\mathcal{K}}\widehat{C}_{\pi(k)}=\widehat{R}_{c}^{\textrm{1-DPCRS}}(\widehat{\mathbf{H}})$ and the total AR of user-$\pi(k)$ is $\widehat{R}_{\pi(k),tot}^{\textrm{1-DPCRS}}(\widehat{\mathbf{H}})\triangleq \widehat{C}_{\pi(k)}+\widehat{R}_{\pi(k)}^{\textrm{1-DPCRS}}(\widehat{\mathbf{H}})$.
The  WASR  maximization problem of 1-DPCRS for each $\widehat{\mathbf{H}}$ with a given weight vector $\mathbf{u}$ and DPC encoding order $\pi$ is:
\begin{subequations}
	\label{eq: DPCRS b}
	\begin{align}
		&	\max_{ \widehat{\mathbf{c}},\mathbf{{P}}}\,\, \sum_{k\in\mathcal{K}}u_{\pi(k)}\widehat{R}_{\pi(k),tot}^{\textrm{1-DPCRS}}(\widehat{\mathbf{H}})\\
	\mbox{s.t.}\,\,
	&\,\, \sum_{k\in\mathcal{K}}\widehat{C}_{k}\leq \widehat{R}_{c}^{\textrm{1-DPCRS}}(\widehat{\mathbf{H}}) \label{c1_DPCRS b}\\
	&  \,\,\widehat{R}_{\pi(k),tot}^{\textrm{1-DPCRS}}(\widehat{\mathbf{H}})\geq R_{\pi(k)}^{th}, \forall k\in\mathcal{K} \label{c2_DPCRS b}\\
	&\,\,	\text{tr}(\mathbf{P}\mathbf{P}^{H})\leq P_{t} \label{c3_DPCRS b}\\	
	&\,\,	\mathbf{\widehat{c}} \geq \mathbf{0}, \label{c4_DPCRS b}
	\end{align}
\end{subequations}
where $\widehat{\mathbf{c}}=[\widehat{C}_1,\ldots,\widehat{C}_K]$ is the AR allocation for the common stream $s_c$ for each $\widehat{\mathbf{H}}$. It is required to be jointly optimized with the precoder so as to maximize the WASR.  Similarly to (\ref{eq: DPC b}), the WASR of all possible DPC encoding orders are required to be considered in order to maximize  the WESR over all channels.

\subsection{M-DPCRS}
\par Following the  methods adopted by DPC and 1-DPCRS, we could also obtain the decomposed WASR maximization problem of the  three-user M-DPCRS to be solved  with a given weight vector $\mathbf{u}$ and DPC encoding order $\pi$ for each $\widehat{\mathbf{H}}$, which is given by
\begin{subequations}
	\label{eq: MDPCRS b}
	\begin{align}
	&	\max_{\pi',\widehat{\mathbf{c}},\mathbf{{P}}}\,\, \sum_{k\in\mathcal{K}}u_{\pi(k)}\widehat{R}_{\pi(k),tot}^{\textrm{M-DPCRS}}(\widehat{\mathbf{H}})\\
	\mbox{s.t.}\,\,
	&\,\, \sum_{k\in\mathcal{K}_i}\widehat{C}_k^{i}\leq \widehat{R}_{i}^{\textrm{M-DPCRS}}(\widehat{\mathbf{H}}), \forall i\in\mathcal{K}_c\\
	&  \,\,\widehat{R}_{\pi(k),tot}^{\textrm{M-DPCRS}}(\widehat{\mathbf{H}})\geq R_{\pi(k)}^{th}, \forall k\in\mathcal{K} \label{c2_MDPCRS b}\\
	&\,\,	\text{tr}(\mathbf{P}\mathbf{P}^{H})\leq P_{t} \label{c3_MDPCRS b}\\	
	&\,\,	\widehat{\mathbf{c}} \geq \mathbf{0}, \label{c4_MDPCRS b}
	\end{align}
\end{subequations}
 where $\widehat{\mathbf{c}}=\{\widehat{C}_k^{i}| k\in\mathcal{K}_i,i\in\mathcal{K}_c\}$ is the AR allocation for all the common streams for each $\widehat{\mathbf{H}}$.   $\widehat{R}_{ \pi(k),tot}^{\textrm{M-DPCRS}}=\sum_{i\in\mathcal{I}_{\pi(k)}\setminus\{\pi(k)\}}\widehat{C}_{\pi(k)}^{i}+\widehat{R}_{\pi(k)}^{\textrm{M-DPCRS}}$ is the total AR at user-$\pi(k)$ to decode $W_{\pi(k)}$ and $\widehat{R}_{i}^{\textrm{M-DPCRS}}(\widehat{\mathbf{H}})=\min\{\widehat{R}_{i,\pi(k)}^{\textrm{M-DPCRS}}(\widehat{\mathbf{H}})\mid \pi(k)\in\mathcal{K}_i\}$ is the AR of $s_i$. Notice that the decoding order $\pi'$ of  the partial-common streams $s_{12}, s_{13}, s_{23}$ is required to be jointly optimized with precoders   for each $\widehat{\mathbf{H}}$  so as to maximize the system WASR for a given weight vector $\mathbf{u}$ and  DPC encoding order $\pi$. To further maximize the WESR for a given set of user weights $\mathbf{u}$, an extra optimization over the DPC encoding order $\pi$ has to be carried out.  Therefore, both the decoding order  $\pi'$ of  the partial-common streams and the DPC encoding order of the private streams $\pi$ are jointly optimized at the BS. This can be done by evaluating the performance for all possible decoding and encoding orders $\pi, \pi'$ and  choosing the one with the highest WASR.

\section{Proposed Optimization Framework}
\label{sec: algorithm}
\par The formulated  problem  (\ref{eq: DPC b}), (\ref{eq: DPCRS b}) and (\ref{eq: MDPCRS b}) are stochastic non-convex optimization problems since the ARs specified in Definition \ref{def: AR} are expectations with respect to the random variable $\widetilde{\mathbf{H}}$. 
To tackle  the stochastic nature and the non-convexity of the  problems, we extend the algorithm proposed in \cite{RS2016hamdi}   to solve the problem. Specifically, there are three steps of the proposed optimization framework:
\begin{itemize}
	\item \textit{Step 1. Sample Average Approximation Approach}: We first employ the Sample Average Approximation (SAA) approach to   transform the original stochastic problems into the corresponding deterministic problems.
	\item \textit{Step 2. Weighted Minimum Mean Square Error Approach}: The non-convexity of each transformed deterministic problem is further tackled by the Weighted Minimum Mean Square Error (WMMSE) approach and  transformed into a block-wise convex problem.
	\item \textit{Step 3. Alternating Optimization Algorithm}: The transformed block-wise convex problem is finally solved by using Alternating Optimization (AO) algorithm.
\end{itemize}
In this section, the above optimization framework of solving M-DPCRS problem (\ref{eq: MDPCRS b}) is specified  followed by the guidance of solving other problems.

\subsection{Sample Average Approximation Approach}
\par  The first step is to use SAA to approximate the stochastic ARs into the corresponding deterministic expressions. As the conditional density $f_{{\mathbf{H}}\mid \widehat{\mathbf{H}}}({\mathbf{H}}\mid \widehat{\mathbf{H}})$ is known at the BS,  for a given channel estimate $\widehat{\mathbf{H}}$,  BS is able to generate a sample of $M$ user channels, indexed by $\mathcal{M}=\{1,\ldots,M\}$ as
\begin{equation}
\mathbb{H}^{(M)}\triangleq \left\{\mathbf{H}^{(m)}=\widehat{\mathbf{H}}+\widetilde{\mathbf{H}}^{(m)}\mid \widehat{\mathbf{H}}, m\in\mathcal{M}\right\}.
	\vspace{-1mm}
\end{equation}
Following the  strong Law of Large Number (LLN), the ARs $	\widehat{R}_{i,k}^{\textrm{x}}(\widehat{\mathbf{H}})$ specified in equation (\ref{eq: average rate})  for decoding  stream $s_i$  at user-$k,k\in\mathcal{K}$ with a given channel estimate $\widehat{\mathbf{H}}$ is equivalent to 
\begin{equation}
\label{eq: SAF}
\begin{aligned}
\widehat{R}_{i,k}^{\textrm{x}}(\widehat{\mathbf{H}})&= \lim\limits_{M\rightarrow \infty} {\widehat{R}_{i,k}^{{\textrm{x}}^{(M)}}}(\widehat{\mathbf{H}}),
\end{aligned}
\end{equation}
where
\begin{equation}
\label{eq: SAF rate}
\begin{aligned}
\widehat{R}_{i,k}^{{\textrm{x}}^{(M)}}(\widehat{\mathbf{H}})\triangleq{\frac{1}{M}\sum_{m=1}^{M}R_{i,k}^{\textrm{x}}\left(\mathbf{H}^{(m)},\widehat{\mathbf{H}}\right)}
\end{aligned}
\end{equation}
is the sampled AR. It approximates AR, i.e., $ \widehat{R}_{i,k}^{\textrm{x}}(\widehat{\mathbf{H}})\approx \widehat{R}_{i,k}^{{\textrm{x}}^{(M)}}(\widehat{\mathbf{H}})$ if $M$ is sufficiently large.
 $\textrm{x}\in$\{``{DPC}",``{1-DPCRS}",``{M-DPCRS}"\}.   $R_{i,k}^{\textrm{x}}(\mathbf{H},\widehat{\mathbf{H}})=R_{k}^{\textrm{x}}(\mathbf{H},\widehat{\mathbf{H}})$ and $\widehat{R}_{i,k}^{\textrm{x}^{(M)}}(\widehat{\mathbf{H}})$ is simplified as $\widehat{R}_{k}^{\textrm{x}^{(M)}}(\widehat{\mathbf{H}})$ when $i=k$. 
Note that the precoder $\mathbf{P}$ in (\ref{eq: SAF}) is unaltered over all the $M$ channel samples. 
Considering M-DPCRS strategy, as  $\pi'$ and $\pi$ are discrete variables, they are optimized at the BS by evaluating the WASR performance for a given pair of $\pi, \pi'$ and  choosing the one with the highest WASR. Hence,  $\widehat{\mathbf{c}}, \mathbf{P}$ are designed by solving  problem (\ref{eq: MDPCRS b}) for each  $\pi'$ and $\pi$ with the average common and private rates approximated by (\ref{eq: SAF rate}), which is given by
\begin{subequations}
	\label{eq: DPCRS SAA}
	\small
	\begin{align}
	&	\max_{ \widehat{\mathbf{c}},\mathbf{{P}}}\,\, \sum_{k\in\mathcal{K}}u_{\pi(k)}\left(\sum_{i\in\mathcal{I}_{\pi(k)}\setminus\{\pi(k)\}}\widehat{C}_{\pi(k)}^{i}+\widehat{R}_{\pi(k)}^{\textrm{M-DPCRS}^{(M)}}(\widehat{\mathbf{H}})\right) \label{obj: WASR}\\
	\mbox{s.t.}\,\,
	&\,\, \sum_{k\in\mathcal{K}_i}\widehat{C}_{k}^i\leq \widehat{R}_{i}^{\textrm{M-DPCRS}^{(M)}}(\widehat{\mathbf{H}}), \forall i\in\mathcal{K}_c \label{c1_DPCRS SAA}\\
	&  \,\,\sum_{i\in\mathcal{I}_{\pi(k)}\setminus\{\pi(k)\}}\widehat{C}_{\pi(k)}^{i}+\widehat{R}_{\pi(k)}^{\textrm{M-DPCRS}^{(M)}}(\widehat{\mathbf{H}})\geq R_{\pi(k)}^{th}, \forall k\in\mathcal{K} \label{c2_DPCRS SAA}\\
	&\,\,\textrm{(\ref{c3_MDPCRS b})},  \textrm{(\ref{c4_MDPCRS b})},\nonumber
	\end{align}
\end{subequations}
where  \resizebox{.43 \textwidth}{!} {$\widehat{R}_{i}^{\textrm{M-DPCRS}^{(M)}}(\widehat{\mathbf{H}})=\min\left\{\widehat{R}_{i,\pi(k)}^{\textrm{M-DPCRS}^{(M)}}(\widehat{\mathbf{H}})\mid \pi(k)\in\mathcal{K}_i\right\}$}. Following the LLN, problem  (\ref{eq: DPCRS SAA}) approximates to the stochastic problem  (\ref{eq: MDPCRS b}) for a given $\pi'$ as $M\rightarrow \infty$. Hence, our target  is transformed to design precoder $\mathbf{P}$ and the common stream allocation vector $\widehat{\mathbf{c}}$ by solving (\ref{eq: DPCRS SAA}).

\subsection{Weighted Minimum Mean Square Error Approach}
\par  Problem (\ref{eq: DPCRS SAA}) is still non-convex due to the non-convex approximated rate expressions of the common stream and the private streams. To solve the problem, we further extend the WMMSE algorithm proposed in \cite{wmmse08, RS2016hamdi}.  At user sides,  user-$\pi(k)$  decodes  data streams $\{s_i| i\in\mathcal{I}_{\pi(k)}\}$  sequentially based on the decoding order $\pi'$ by employing the equalizer $g_{\pi(k)}^i, i\in\mathcal{I}_{\pi(k)}$. The signal received at user-$\pi(k)$ is $y_{\pi(k)}=\mathbf{h}_{\pi(k)}^H\mathbf{x}+n_{\pi(k)}$, where
$\mathbf{x}$ is specified in equation (\ref{eq: MDPCRS transmitSig}).
$s_{123}$ is decoded first and the estimated common stream $\widehat{s}_{123}$ is $\widehat{s}_{123}=g_{\pi(k)}^{123}y_{\pi(k)}$.  Once $s_{123}$ is successfully decoded and removed from the received signal,  user-$\pi(k)$ then decodes the partial-common streams  $s_{\pi'_{\pi(k)}(1)}$, $s_{\pi'_{\pi(k)}(2)}$ by employing the equalizer $g_{\pi(k)}^{i}, i\in\{\pi'_{\pi(k)}\}$ followed by the private stream  $s_{\pi(k)}$ via equalizer $g_{\pi(k)}^{\pi(k)}$.
The Mean Square Error (MSE) of each stream $s_i, i\in\mathcal{I}_{\pi(k)}$ at user-$\pi(k)$ is
\begin{equation}
\label{eq:MSE}
\begin{aligned}
\resizebox{.46 \textwidth}{!} {$\varepsilon_{\pi(k)}^i\triangleq\mathbb{E}\{|\widehat{s}_{i}-s_{i}|^{2}\}=|g_{\pi(k)}^i|^2T_{\pi(k)}^i-2\Re\{g_{\pi(k)}^i\mathbf{h}_{\pi(k)}^H\mathbf{p}_{i}\}+1,$}
\end{aligned}
\end{equation}
where $T_{\pi(k)}^{i}$ is the remaining received power at user-$\pi(k)$ when decoding $s_i$. Mathematically, it is defined as $T_{\pi(k)}^{i}\triangleq \sum_{j\in\bar{\mathcal{K}}_{c}^i\cup\mathcal{K}\cup\{i\}}|\mathbf{h}_{\pi(k)}^H\mathbf{p}_{j}|^2+1, \forall i\in \{123, \pi'_{\pi(k)}\}$ and 
$T_{\pi(k)}^{i}\triangleq \sum_{j\in\bar{\mathcal{K}}_{\pi(k)}\cup\{i\}}|\mathbf{h}_{\pi(k)}^H\mathbf{p}_{j}|^2$ $+\sum_{j<k}|\widetilde{\mathbf{h}}_{\pi(k)}^H\mathbf{p}_{\pi(j)}|^2+1,  \textrm{if } i=\pi(k)$.

\par Define the Weighted MSE (WMSE) of decoding $s_i$ at user-$\pi(k)$ as
\begin{equation}
\xi_{\pi(k)}^i( \mathbf{H},\widehat{\mathbf{H}})\triangleq w_{\pi(k)}^i\varepsilon_{\pi(k)}^i-\log_{2}(w_{\pi(k)}^i), 
\end{equation}
where $w_{\pi(k)}^i$ is the introduced weight for MSE of user-$\pi(k)$.
The corresponding Weighted Minimum MSE (WMMSE) metrics of the common and private streams are 
\begin{equation}
\begin{aligned}
\label{eq: WMMSE}
&\xi_{i,\pi(k)}^{\textrm{MMSE}}( \mathbf{H},\widehat{\mathbf{H}})\triangleq\min_{w_{\pi(k)}^i,g_{\pi(k)}^i}\xi_{\pi(k)}^i( \mathbf{H},\widehat{\mathbf{H}}).
\end{aligned}
\end{equation}
With the introduced WMMSEs, we obtain the following proposition.
\begin{proposition}
	\label{prop: WMMSE-rate}
	The instantaneous rate and the WMMSE of decoding stream $s_i$  at user-$\pi(k),k\in\mathcal{K}$ has the following relationship:
	\begin{equation}
	\label{eq: rate-wmmse}
	\xi_{i,\pi(k)}^{\textrm{MMSE}}( \mathbf{H},\widehat{\mathbf{H}})=1-R_{i,\pi(k)}^{\textrm{M-DPCRS}}( \mathbf{H},\widehat{\mathbf{H}}), 
	\end{equation}
	where $R_{i,\pi(k)}^{\textrm{M-DPCRS}}( \mathbf{H},\widehat{\mathbf{H}})=R_{\pi(k)}^{\textrm{M-DPCRS}}( \mathbf{H},\widehat{\mathbf{H}})$ when $i=\pi(k)$.
\end{proposition}

\par \textit{Proof:} Following (\ref{eq: WMMSE}), the  WMSE weights ($w_{\pi(k)}^{i,\star}$) and equalizers ($g_{\pi(k)}^{i,\star}$) of minimizing  $\xi_{\pi(k)}^i( \mathbf{H},\widehat{\mathbf{H}})$  satisfy that  $\left.\frac{\partial\xi_{\pi(k)}^i( \mathbf{H},\widehat{\mathbf{H}})}{\partial w_{\pi(k)}^i}\right|_{\small(w_{\pi(k)}^i,g_{\pi(k)}^i)=(w_{\pi(k)}^{i,\star},g_{\pi(k)}^{i,\star})}=0$  and  $\left.\frac{\partial\xi_{\pi(k)}^i( \mathbf{H},\widehat{\mathbf{H}})}{\partial g_{\pi(k)}^i}\right|_{\small(w_{\pi(k)}^i,g_{\pi(k)}^i)=(w_{\pi(k)}^{i,\star},g_{\pi(k)}^{i,\star})}=0$.  We first solve $\frac{\partial\xi_{\pi(k)}^i( \mathbf{H},\widehat{\mathbf{H}})}{\partial g_{\pi(k)}^i}=0$ and obtain the    equalizer as
\begin{equation}
\label{eq: WMMSE equalizers}
	g_{\pi(k)}^{i,\star}=\mathbf{p}_{i}^H\mathbf{h}_{\pi(k)}({T}_{\pi(k)}^i)^{-1}.
\end{equation} 
By further solving   $\left.\frac{\partial\xi_{\pi(k)}^i( \mathbf{H},\widehat{\mathbf{H}})}{\partial w_{\pi(k)}^i}\right|_{g_{c,\pi(k)}=g_{\pi(k)}^{i,\star}}=0$, we obtain that 
\begin{equation}
\label{eq: WMMSE weights}
w_{\pi(k)}^{i,\star}=\frac{1}{\varepsilon_{\pi(k)}(g_{\pi(k)}^{i,\star})}=\frac{T_{\pi(k)}^i}{T_{\pi(k)}^i-|\mathbf{h}_{\pi(k)}^H\mathbf{p}_{i}|^2}.
\end{equation} 
Substituting ($w_{\pi(k)}^{i,\star},g_{\pi(k)}^{i,\star}$) back to $\xi_{\pi(k)}^i( \mathbf{H},\widehat{\mathbf{H}})$,   $\xi_{\pi(k)}^{i,\textrm{MMSE}}( \mathbf{H},\widehat{\mathbf{H}})$ is derived as
\begin{equation}
\xi_{i,\pi(k)}^{\textrm{MMSE}}( \mathbf{H},\widehat{\mathbf{H}})=\log_2(w_{\pi(k)}^{i,\star})=1-R_{i,\pi(k)}^{\textrm{M-DPCRS}}( \mathbf{H},\widehat{\mathbf{H}}).
\end{equation}
The proof is completed.\qed

\par  The Rate-WMMSE relationships in (\ref{eq: rate-wmmse}) is established for instantaneous channel realizations. We can also extend it to the average Rate-WMMSE relationships over a sample of $M$ user channels as
\begin{equation}
\label{eq: rate-wmmse average}
\begin{aligned}
\widehat{\xi}_{i,\pi(k)}^{^{(M)}}(\widehat{\mathbf{H}})&\triangleq{{\frac{1}{M}\sum_{m=1}^{M}\left(\min_{w_{\pi(k)}^{i,(m)},g_{\pi(k)}^{i,(m)}}\xi_{\pi(k)}^i( \mathbf{H}^{(m)},\widehat{\mathbf{H}})\right)}}\\
&=1-{\widehat{R}_{i,\pi(k)}^{\textrm{M-DPCRS}^{(M)}}}(\widehat{\mathbf{H}}),
\end{aligned}
\end{equation}
where $w_{\pi(k)}^{i,(m)},g_{\pi(k)}^{i,(m)}$ are the weights and equalizers associated with the $m$th channel realization in $\mathbb{H}^{(M)}$.   ${\widehat{\xi}_{i,\pi(k)}^{(M)}}( \widehat{\mathbf{H}})=\widehat{\xi}_{\pi(k)}^{{(M)}}( \widehat{\mathbf{H}})$ when $i=\pi(k)$.
With the average Rate-WMMSE relationships in (\ref{eq: rate-wmmse average}), problem (\ref{eq: DPCRS SAA}) is equivalently transformed   into the WMMSE problem

\begin{subequations}
	\label{eq: RS WMMSE}
	\small
	\begin{align}
	&	\min_{ \mathbf{{P}},\widehat{\mathbf{x}},\mathbf{w},\mathbf{g}}\,\, \sum_{k\in\mathcal{K}}u_{\pi(k)}\left(\sum_{i\in\mathcal{I}_{\pi(k)}\setminus\{\pi(k)\}}\widehat{X}_{\pi(k)}^{i}+\widehat{\xi}_{\pi(k)}^{{(M)}}( \widehat{\mathbf{H}})\right) \label{obj: WMMSE}\\
	\mbox{s.t.}\,\,
	&\,\, \sum_{k\in\mathcal{K}}\widehat{X}_{k}^i+1\geq \widehat{\xi}_{i}^{{(M)}}( \widehat{\mathbf{H}}), \forall i\in\mathcal{K}_c \label{c1_RS WMMSE}\\
	&  \,\,\sum_{i\in\mathcal{I}_{\pi(k)}\setminus\{\pi(k)\}}\widehat{X}_{\pi(k)}^{i}+\widehat{\xi}_{\pi(k)}^{{(M)}}( \widehat{\mathbf{H}})\leq 1-R_{\pi(k)}^{th}, \forall k\in\mathcal{K} \label{c2_RS WMMSE}\\
	&\,\,		\text{tr}(\mathbf{P}\mathbf{P}^{H})\leq P_{t}  \label{c4_RS WMMSE}\\
	&\,\,\widehat{\mathbf{x}} \leq \mathbf{0},  \label{c3_RS WMMSE} 
	\end{align}
\end{subequations}
where $\widehat{\mathbf{x}}=\{\widehat{X}_k^{i}| k\in\mathcal{K}_i,i\in\mathcal{K}_c\}$ is the transformation of the common rate $\widehat{\mathbf{c}}$ with $\widehat{\mathbf{x}}=-\widehat{\mathbf{c}}$ holds.   $\mathbf{w}=\{w_{\pi(k)}^{i,(m)} |i\in\mathcal{I}_{\pi(k)}, k\in\mathcal{K},m\in\mathcal{M}\}$  and $\mathbf{g}=\{g_{\pi(k)}^{i,(m)} | i\in\mathcal{I}_{\pi(k)}, k\in\mathcal{K},m\in\mathcal{M}\}$ are the MSE weights and equalizers, respectively.  $\widehat{\xi}_{i}^{{(M)}}( \widehat{\mathbf{H}})=\max\{\widehat{\xi}_{i,\pi(k)}^{(M)}(\widehat{\mathbf{H}})\mid \pi(k)\in\mathcal{K}_i\}$.

\subsection{Alternating Optimization Algorithm}
\par Though  problem (\ref{eq: RS WMMSE}) that jointly optimizes $(\mathbf{{P}},\widehat{\mathbf{x}},\mathbf{w},\mathbf{g})$ is still non-convex, it is block-wise convex with respect to each block of $\mathbf{w}$, $\mathbf{g}$ and $(\mathbf{{P}},\widehat{\mathbf{x}})$ by fixing other two blocks\footnote{The block-wise convexity of problem (\ref{eq: RS WMMSE})  can be observed by treating  one block among $\mathbf{w}$, $\mathbf{g}$ and $(\mathbf{{P}},\widehat{\mathbf{x}})$ as optimization variables and other two blocks as constants according to the concept of convex optimization problems specified in  \cite[Section 4]{boyd2004convex}.  }. This motivates us to use AO algorithm to solve the problem.  At each iteration $[n]$, for  given $\mathbf{w}^{[n-1]}$ and $(\mathbf{{P}}^{[n-1]},\widehat{\mathbf{x}}^{[n-1]})$, the  solution $\mathbf{g}^{\star}$ of   (\ref{eq: RS WMMSE}) is
\begin{equation}
\label{eq: equalizer update}
\mathbf{g}^{[n]}\triangleq\mathbf{g}^{\star}(\mathbf{{P}}^{[n-1]})=\{g_{\pi(k)}^{i,\star,(m)}\mid i\in\mathcal{I}_{\pi(k)}, k\in\mathcal{K},m\in\mathcal{M}\}
\end{equation}    
with each element calculated by equation (\ref{eq: WMMSE equalizers}) and precoder $\mathbf{{P}}^{[n-1]}$ for the $m$th channel realization in $\mathbb{H}^{(M)}$.  For  given $\mathbf{g}^{[n-1]}$ and $(\mathbf{{P}}^{[n-1]},\widehat{\mathbf{x}}^{[n-1]})$, the  solution $\mathbf{w}^{\star}$ of   (\ref{eq: RS WMMSE}) is 
\begin{equation}
\label{eq: weight update}
\mathbf{w}^{[n]}\triangleq\mathbf{w}^{\star}(\mathbf{{P}}^{[n-1]})=\{w_{ \pi(k)}^{i,\star,(m)}\mid i\in\mathcal{I}_{\pi(k)}, k\in\mathcal{K},m\in\mathcal{M}\}
\end{equation}  
with each element calculated by equation (\ref{eq: WMMSE weights}) and precoder $\mathbf{{P}}^{[n-1]}$. The solutions of weights and equalizers can be verified through showing that $\left(\mathbf{w}^{[n]}, \mathbf{g}^{[n]}\right)$ satisfy the   Karush-Kuhn-Tucker (KKT) conditions of  (\ref{eq: RS WMMSE}). Substituting $\left(\mathbf{w}^{[n]}, \mathbf{g}^{[n]}\right)$ back to   (\ref{eq: RS WMMSE}), the optimization problem is equivalently transformed as:
\begin{subequations}
	\label{eq: RS WMMSE final}
		\small
	\begin{align}
	&	\min_{\mathbf{{P}},\widehat{\mathbf{x}}}\,\, \sum_{k\in\mathcal{K}}u_{\pi(k)}\left(\sum_{i\in\mathcal{I}_{\pi(k)}\setminus\{\pi(k)\}}\widehat{X}_{\pi(k)}^{i}+\widehat{\xi}_{\pi(k)}^{\textrm{M-DPCRS}}(\widehat{\mathbf{H}})\right)\\
	\mbox{s.t.}\,\,
	&\,\, \sum_{k\in\mathcal{K}}\widehat{X}_{k}^i+1\geq \widehat{\xi}_{i}^{\textrm{M-DPCRS}}(\widehat{\mathbf{H}}), \forall i\in\mathcal{K}_c \label{c1_RS WMMSE final}\\
	&  \,\,\sum_{i\in\mathcal{I}_{\pi(k)}\setminus\{\pi(k)\}}\widehat{X}_{\pi(k)}^{i}+\widehat{\xi}_{\pi(k)}^{\textrm{M-DPCRS}}( \widehat{\mathbf{H}})\leq 1-R_{\pi(k)}^{th}, \forall k\in\mathcal{K} \label{c2_RS WMMSE final}\\
	&\,\,\textrm{(\ref{c4_RS WMMSE})},  \textrm{(\ref{c3_RS WMMSE})},\nonumber
	\end{align}
\end{subequations}
	where  
\begin{equation}
\label{eq: qcqp xi}
\resizebox{.43\textwidth}{!} {$	\widehat{\xi}_{i,\pi(k)}^{\textrm{M-DPCRS}}( \widehat{\mathbf{H}})\triangleq\Omega_{\pi(k)}^{i}+\bar{t}_{\pi(k)}^i-2\Re\left\{(\bar{\mathbf{f}}_{\pi(k)}^i)^H\mathbf{p}_{i}\right\}+\bar{w}_{\pi(k)}^i-\bar{\nu}_{{\pi(k)}}^i$}
\end{equation}
and $\widehat{\xi}_{i,\pi(k)}^{\textrm{M-DPCRS}}( \widehat{\mathbf{H}})$ is simplified to $\widehat{\xi}_{\pi(k)}^{\textrm{M-DPCRS}}( \widehat{\mathbf{H}})$ when $i=\pi(k)$. $\Omega_{\pi(k)}^{i}$ in (\ref{eq: qcqp xi}) is defined as
\begin{equation}
\begin{aligned}
	&\Omega_{\pi(k)}^{i}\triangleq \\
&\resizebox{.48\textwidth}{!} {$
	\begin{cases}
	\sum_{j\in\bar{\mathcal{K}}_{c}^i\cup\mathcal{K}\cup\{i\}}\mathbf{p}_{j}^H\bar{\Psi}_{\pi(k)}^i\mathbf{p}_{ j}, \quad   \forall i\in \{123, \pi'_{\pi(k)}\} \\ 
	\sum_{j\in\bar{\mathcal{K}}_{\pi(k)}\cup\{i\}}\mathbf{p}_{j}^H\bar{\Psi}_{\pi(k)}^i\mathbf{p}_{ j}+\sum_{j< k}\mathbf{p}_{\pi(j)}^H\bar{\Phi}_{\pi(k)}^i\mathbf{p}_{\pi(j)},
\,\, i=\pi(k),
	\end{cases}$}
\end{aligned}
\end{equation}
and $\bar{\Psi}_{\pi(k)}^i, $ $ \bar{\Phi}_{\pi(k)}^i, \bar{t}_{\pi(k)}^i,  \bar{\mathbf{f}}_{\pi(k)}^i, \bar{w}_{\pi(k)}^i,  \bar{\nu}_{{\pi(k)}}^i$ are constants (or constant vectors/matrices) averaged over a sample of $M$ user channels, i.e., $\bar{w}_{\pi(k)}=\frac{1}{M}\sum_{m=1}^M{w}_{\pi(k)}^{i,(m)}$. Their corresponding values in each channel instance $(m)$ are updated as
\begin{equation}
	\small
\begin{aligned}
& {t}_{\pi(k)}^{i,(m)}=w_{\pi(k)}^{i,(m)}\left|g_{\pi(k)}^{i,(m)}\right|^2,\\
&  {\Psi}_{\pi(k)}^{i,(m)}={t}_{\pi(k)}^{i,(m)}\mathbf{h}_{\pi(k)}^{(m)}(\mathbf{h}_{\pi(k)}^{(m)})^H,\\ &{\Phi}_{\pi(k)}^{i,(m)}={t}_{\pi(k)}^{i,(m)}\widetilde{\mathbf{h}}_{\pi(k)}^{(m)}(\widetilde{\mathbf{h}}_{\pi(k)}^{(m)})^H,\\
&\mathbf{f}_{\pi(k)}^{i,(m)}=w_{\pi(k)}^{i,(m)}\mathbf{h}_{\pi(k)}^{(m)}(g_{\pi(k)}^{i,(m)})^H,\\
& {\nu}_{{\pi(k)}}^{i,(m)}=\log_2\left(w_{\pi(k)}^{i,(m)}\right).
\end{aligned}
\end{equation}

\par As  $\widehat{\xi}_{\pi(k)}^{\textrm{M-DPCRS}}( \widehat{\mathbf{H}}), \widehat{\xi}_{i}^{\textrm{M-DPCRS}}( \widehat{\mathbf{H}})$  are quadratic according to (\ref{eq: qcqp xi}),  problem (\ref{eq: RS WMMSE final})  is a convex Quadratically Constrained Quadratic Program (QCQP), which can be solved via interior-point methods \cite{boyd2004convex}.  Therefore,  $(\mathbf{{P}}^{[n]},\widehat{\mathbf{x}}^{[n]})$ can  be updated by using the optimal solution of  (\ref{eq: RS WMMSE final}). The details to the proposed AO algorithm is specified in Algorithm \ref{WMMSE algorithm}. The weights $\mathbf{w}$, equalizers $\mathbf{g}$, precoders and common rate vectors $(\mathbf{P},\widehat{\mathbf{x}})$ are updated iteratively until  the WASR of the system  $\mathrm{WASR}^{[n]}$ (calculated by  (\ref{obj: WASR}) based on the solution $\mathbf{P}^{[n]}, \widehat{\mathbf{c}}^{[n]}=-\widehat{\mathbf{x}}^{[n]}$)  converges. 

\begin{algorithm}[h!]
	\textbf{Initialize}: $n\leftarrow0$, $\mathbf{P}$, $\mathrm{WASR}^{[n]}$\;
	\Repeat{$|\mathrm{WASR}^{[n]}-\mathrm{WASR}^{[n-1]}|\leq \epsilon$}{
		$n\leftarrow n+1$\;
		$\mathbf{P}^{[n-1]}\leftarrow \mathbf{P}$\;
		update $\mathbf{g}$ and $\mathbf{w}$  by $\mathbf{g}^{\star}(\mathbf{P}^{[n-1]})$ and $\mathbf{w}^{\star}(\mathbf{P}^{[n-1]})$ specified in (\ref{eq: equalizer update}) and (\ref{eq: weight update}), respectively\; 
		update $(\mathbf{P},\widehat{\mathbf{x}})$ by solving (\ref{eq: RS WMMSE final}) using the updated $\mathbf{w}, \mathbf{g}$;	
	}	
	\caption{WMMSE-based AO algorithm}
	\label{WMMSE algorithm}				
\end{algorithm}

\subsection{Convergence}
\par The convergence of Algorithm \ref{WMMSE algorithm} is guaranteed according to Proposition \ref{prop: converge}.
\begin{proposition}
	\label{prop: converge}
	Denote any stationary point of problem (\ref{eq: MDPCRS b}) for a given common stream decoding order $\pi'$ as ($\mathbf{P}^{\circ}, \widehat{\mathbf{c}}^{\circ}$).
	With a feasible initial point, the proposed Algorithm \ref{WMMSE algorithm}  is guaranteed to converge. As $M\rightarrow \infty$, the convergent solution ($\mathbf{P}', \widehat{\mathbf{x}}'$) of  Algorithm \ref{WMMSE algorithm} is a stationary point of problem (\ref{eq: MDPCRS b})  with $\mathbf{P}'=\mathbf{P}^{\circ}$ and $\widehat{\mathbf{x}}'=-\widehat{\mathbf{c}}^{\circ}$ holds. 
\end{proposition}

\par \textit{Proof:} The proof in \cite{RS2016hamdi} for the AO algorithm of linearly precoded 1-layer  RS  is extended here for that of   M-DPCRS.  We first show that the proposed algorithm is guaranteed to converge followed by showing the convergent point is a stationary point of problem (\ref{eq: DPCRS SAA}) and problem (\ref{eq: MDPCRS b}) as $M\rightarrow \infty$.

\par As the solution $\mathbf{P}^{[n]}, \widehat{\mathbf{x}}^{[n]}, \mathbf{w}^{[n]}, \mathbf{g}^{[n]}$  of problem (\ref{eq: RS WMMSE final}) at iteration $[n]$ is also a feasible solution of  (\ref{eq: RS WMMSE final}) at iteration $[n+1]$, the corresponding objective function of problem (\ref{eq: RS WMMSE final}) is guaranteed to decrease monotonically. Due to the transmit power constraint (\ref{c4_RS WMMSE}), the objective function of (\ref{eq: RS WMMSE final}) is bounded below. Therefore, the AO algorithm proposed to solve problem (\ref{eq: RS WMMSE final}) is guaranteed to converge.

\par Next, we show that the solution sequence  $\{\mathbf{P}^{[n]}, \widehat{\mathbf{x}}^{[n]}\}_{n=1}^{\infty}$ of problem (\ref{eq: RS WMMSE final}) 
converges to  a stationary point ($\mathbf{P}^{\circ}, \widehat{\mathbf{c}}^{\circ}$) of problem (\ref{eq: MDPCRS b}) with $\mathbf{P}'=\mathbf{P}^{\circ}$ and $\widehat{\mathbf{x}}'=-\widehat{\mathbf{c}}^{\circ}$ holds.   
Following  Proposition \ref{prop: WMMSE-rate}, we obtain that problem (\ref{eq: RS WMMSE final}) is a convex approximation of  problem (\ref{eq: DPCRS SAA}) around the point $(\mathbf{{P}}^{[n-1]},\widehat{\mathbf{x}}^{[n-1]})$.
As the AO algorithm is also a special instance of the Successive Convex Approximation (SCA) method  \cite{RS2016hamdi}, the KKT conditions of the original problem at the point $(\mathbf{{P}}^{[n-1]},\widehat{\mathbf{x}}^{[n-1]})$ is maintained. Hence, we obtain that problem (\ref{eq: DPCRS SAA}) and problem (\ref{eq: RS WMMSE final}) share the same sets of KKT points (stationary points) when the solution point $(\mathbf{{P}}^{[n]},\widehat{\mathbf{x}}^{[n]})$ is the same as the solution of the previous iteration $(\mathbf{{P}}^{[n-1]},\widehat{\mathbf{x}}^{[n-1]})$, which is  the convergent point  ($\mathbf{P}', \widehat{\mathbf{x}}'$). 
Since Problem  (\ref{eq: DPCRS SAA}) is equivalent to problem (\ref{eq: MDPCRS b}) for a given $\pi'$ as $M\rightarrow \infty$, we obtain that the convergent point  ($\mathbf{P}', \widehat{\mathbf{x}}'$) of the proposed AO algorithm is 
 one stationary point ($\mathbf{P}^{\circ}, \widehat{\mathbf{c}}^{\circ}$) of problem (\ref{eq: MDPCRS b}) with $\mathbf{P}'=\mathbf{P}^{\circ}$ and $\widehat{\mathbf{x}}'=-\widehat{\mathbf{c}}^{\circ}$ holds.    \qed

\subsection{Complexity}
\begin{table}[t!]
	\centering
	\caption{Computational complexity  of  Algorithm 1 for different strategies}
	\label{tab: computational complexity}
		\begin{tabular}{|l|l|}
			\hline
			\textbf{Strategy}                                      & \textbf{Computational complexity of Algorithm 1 }                                                   \\ \hline
			\textbf{DPC}                                         & \resizebox{.26\textwidth}{!} {$\mathcal{O}\left(([KN_t]^{3.5}+K^2N_tM)K!\log(\epsilon^{-1})\right)$  }                                 \\ \hline
			\textbf{1-DPCRS}                                   & \resizebox{.26\textwidth}{!} {$\mathcal{O}\left(([KN_t]^{3.5}+K^2N_tM)K!\log(\epsilon^{-1})\right)$  }                                 \\ \hline
			\textbf{M-DPCRS}                               & \resizebox{.36\textwidth}{!} {$\mathcal{O}\left(([2^KN_t]^{3.5}+2^{2K}KN_tM)K!\prod_{k=2}^{K-1} \binom{K}{k}!\log(\epsilon^{-1}) \right)$} \\ \hline
	\end{tabular}
\end{table}
\par  The computational complexity of using Algorithm \ref{WMMSE algorithm} to solve the problems of DPC and DPCRS-based strategies are summarized in Table \ref{tab: computational complexity} under the assumption that $N_t\geq K$.

\par	At each iteration of Algorithm \ref{WMMSE algorithm}, the MMSE equalizers and weights $(\mathbf{w},\mathbf{g})$ are updated with complexity $\mathcal{O}(K^2N_tM)$ for DPC and 1-DPCRS, or   $\mathcal{O}(2^{2K}KN_tM)$ for M-DPCRS.  The precoders and common rate vector $(\mathbf{P},\mathbf{x})$ are then updated by using interior-point method with computational complexity  $\mathcal{O}([KN_t]^{3.5})$ for DPC and 1-DPCRS, or   $\mathcal{O}([2^{K}N_t]^{3.5})$ for M-DPCRS \cite{ye1997interior}. The AO algorithm  with  convergence tolerance $\epsilon$ requires $\mathcal{O}(\log(\epsilon^{-1}))$ iterations to converge. Note that  the optimal DPC encoding order in MISO BC with partial CSIT is unknown. Therefore, Algorithm \ref{WMMSE algorithm}  is required to be repeated for all possible DPC encoding orders. The number of possible DPC encoding orders is $K!$ in the $K$-user case. For M-DPCRS, Algorithm \ref{WMMSE algorithm} is  also required to be repeated for all possible decoding orders of the common streams. In the $K$-user case, the number of possible decoding order of the common streams is $\prod_{k=2}^{K-1} \binom{K}{k}!$  \cite{mao2019TCOM}. 

\par Compared with DPC and 1-DPCRS, the computational complexity of M-DPCRS is much higher due to the exponentially increasing number of common streams as well as the joint optimization of  the precoders and the decoding order of common streams. 
As discussed in \cite{mao2019TCOM}, the generalized RS has a number of common streams (and therefore complexity) that increases exponentially with $K$, which suggests that the generalized RS is applicable to only relatively small $K$ scenarios. M-DPCRS proposed here incurs the same complexity issue.  Moreover, M-DPCRS  is considered as a novel technique to exploit larger ergodic rate region in MISO BC with partial CSIT. We show in Section \ref{sec: numerical results} that linearly precoded RS  has a much lower complexity and is able to achieve very close performance to M-DPCRS.

\begin{remark}Similarly, problem  (\ref{eq: DPC b}) and (\ref{eq: DPCRS b})   are solved respectively by approximating each stochastic optimization problem using the SAA approach. The equivalently transformed problems are then reformulated into the  WMMSE problems and solved by the corresponding AO algorithm.
\end{remark}

\section{Numerical Results}
\label{sec: numerical results}
\par In this section, the WSR performance of the proposed 1-DPCRS and M-DPCRS strategies in  MISO BC with partial CSIT  is evaluated. In the following numerical results, all the optimization problems to be solved by using  interior-point methods  are solved using the CVX toolbox \cite{grant2008cvx}. User channels are randomly generated as specified in \cite{RS2016hamdi,hamdi2016robust,mao2019TCOM}.  The actual  user channel $\mathbf{h}_k$ experienced at user-$k$   has i.i.d. complex Gaussian entries  drawn from the distribution $\mathcal{CN}(0,\sigma_k^2)$ and the channel estimation error $\widetilde{\mathbf{h}}_k$ has i.i.d. complex Gaussian entries drawn from distribution $\mathcal{CN}(0,\sigma_{e,k}^2)$. The variance of $\widetilde{\mathbf{h}}_k$ is defined as $\sigma_{e,k}^2\triangleq\sigma_k^2P_t^{-\alpha}$.  As user channels with heterogeneous variances  are considered, the correspoding CSIT qualities also scale with the channel variance $\sigma_k^2$. $\alpha\in[0,1]$ represents SNR scaling as described in Section \ref{sec: channelModel}. We obtain that $\widehat{\mathbf{h}}_k={\mathbf{h}}_k-\widetilde{\mathbf{h}}_k$ also follows Gaussian distribution $\mathcal{CN}(0,\sigma_k^2-\sigma_{e,k}^2)$. The WESR is obtained by averaging the WASR over 100 channel realizations.  For each $\widehat{\mathbf{H}}$, the AR of each user is approximated using SAA method over $M=1000$ samples of user channels  $\mathbb{H}^{(M)}$. For a given $\widehat{\mathbf{H}}$, the $m$th channel estimation error $\widetilde{\mathbf{H}}^{(m)}$ is randomly generated from the error distribution. Hence,  the $m$th conditional channel  is calculated by $\mathbf{H}^{(m)}=\widehat{\mathbf{H}}+\widetilde{\mathbf{H}}^{(m)}$. The initialization of the precoders $\mathbf{P}$ for Algorithm \ref{WMMSE algorithm} is designed by the Maximum Ratio Transmission (MRT) and Singular Value Decomposition (SVD) method proposed in \cite{RS2016hamdi}. The precoders for the private streams of RS-assisted or  other non-RS-assisted transmission strategies are initialized by MRT, i.e., $\mathbf{p}_k=\sqrt{p_k}\frac{\mathbf{h}_k}{\left \| \mathbf{h}_k \right \|}$. The precoders for the common streams of RS-assisted strategies are initialized by SVD, i.e., for  1-DPCRS  and 1-layer RS,  $\mathbf{p}_{c}=\sqrt{p_c}\widehat{\mathbf{p}}_c$, where  $\widehat{\mathbf{p}}_c$ is the largest left singular vector of the channel estimate $\widehat{\mathbf{H}}$. $p_c$ and $p_k$ are the power allocated to each precoder, it follows that $p_c+\sum_{k\in\mathcal{K}}p_k=P_t$. The precoders of the  common streams $s_i, i\in\mathcal{K}_c$ for M-DPCRS and the generalized RS  are initialized in the same way as $\mathbf{p}_c$ but $\widehat{\mathbf{p}}_i$ is chosen based on  $\widehat{\mathbf{H}}_i$ formed by the channel  estimate of users in $\mathcal{K}_i$.

\par The following eight transmission strategies are compared:
\begin{itemize}
	\item \textbf{M-DPCRS}: the M-DPCRS strategy proposed in Section \ref{sec: DPCRS}. In the $K$-user case, there are $2^K-1$ linearly precoded common streams and $K$ DPC-coded private streams to be transmitted from the BS.
	\item \textbf{1-DPCRS}: the  1-DPCRS  strategy proposed in Section \ref{sec: DPCRS}. One linearly precoded common stream and $K$ DPC-coded private streams are transmitted in the $K$-user case.
	\item \textbf{DPC}: the conventional DPC strategy specified in Section \ref{sec: DPC}.   There are $K$ DPC-coded data streams to be transmitted in the $K$-user case. 
	\item \textbf{generalized RS}: the multi-layer RS strategy proposed in \cite{mao2017rate}. User messages are split in the same way as M-DPCRS discussed in Section \ref{sec: DPCRS}. The main difference compared with M-DPCRS is the private streams of the generalized RS are linearly precoded. In the $K$-user case, there are $2^K-1$ linearly precoded common streams and $K$ linearly precoded private streams to be transmitted from the BS.
	\item \textbf{1-layer RS}: the 1-layer RS strategy specified in \cite{RS2015bruno,RS2016hamdi,RSintro16bruno,mao2017rate}. The message of each user  is split into a common part and a private part.  There is one linearly precoded common stream and $K$ linearly precoded private streams   to be transmitted jointly from the BS. Each user is required to  decode the common stream first and uses one layer of SIC to remove the common stream before decoding the intended private stream.
	\item \textbf{SC--SIC}: the power-domain NOMA widely studied in the literature \cite{gc01}.  In the $K$-user case, the streams are linearly precoded and superimposed at the BS before transmission. Users are ordered based on their effective scalar channel strength after precoding. Each user is required to  decode and remove the interference from users with weaker effective channel strength sequentially using SIC. 
	\item \textbf{SC--SIC per group}: the method of combining SDMA and NOMA in MIMO transmission networks \cite{NOMA2017Choi}. The $K$ users are clustered into multiple groups. The inner-group interference is coordinated  by SC--SIC while the inter-group interference is coordinated by SDMA. At the BS,  the $K$-user messages are linearly precoded. Users within the same group are ordered by the corresponding effective channel strength such that each user is able to sequentially decode and remove the interference from weaker users within the same group. The interference from users in different groups is fully treated as noise at each user. 
	\item \textbf{MU--LP}:	Multi-User Linear Precoding (MU--LP) is a practical transmission strategy that has been widely studied in MIMO networks and it is the common implementation of SDMA.	User messages are linearly precoded and superimposed at the BS and each user directly decodes its intended data stream by fully treating any residual interference as noise. 
\end{itemize}
Readers are referred to \cite{mao2017rate} for more details of ``generalized RS", ``1-layer RS", ``SC--SIC", ``SC--SIC per group" and ``MU--LP" transmission strategies, where the corresponding WSR maximization problems are studied. In the sequel, we evaluate the WESR performance of all the eight strategies in a wide range of user deployments considering a diverse range of CSIT qualities, QoS rate requirements and channel strength disparities among users\footnote{As we use random channel realizations, the channel strength disparities are manifested by tuning the channel variance $\sigma_k^2$ of each user. It is termed ``channel variance disparities" in the following.}.    In the following, we first illustrate the results in the two-user MISO BC with partial CSIT followed by the three-user case. 
\subsection{Two-user case}
\label{sec: simulation two user}
\begin{figure}
	\centering
	\begin{subfigure}[b]{0.23\textwidth}
		\centering
		\includegraphics[width=\textwidth]{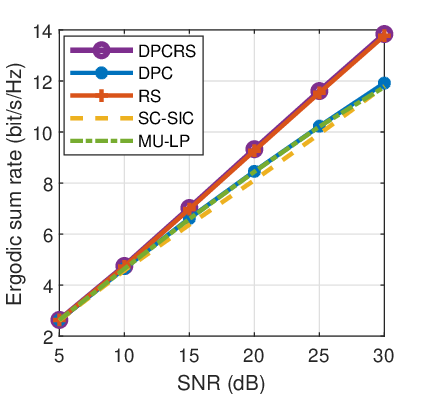}%
		\caption{$\alpha=0.3$}
	\end{subfigure}%
	~
	\begin{subfigure}[b]{0.23\textwidth}
		\centering
		\includegraphics[width=\textwidth]{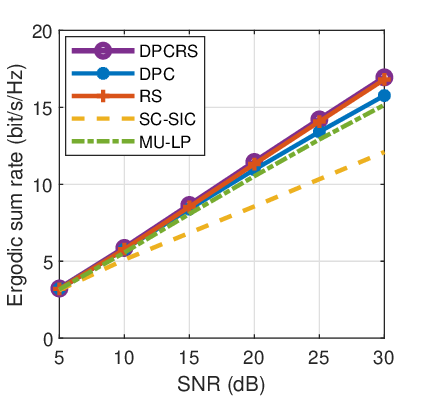}%
		\caption{$\alpha=0.6$}
	\end{subfigure}%
	~\\
	\begin{subfigure}[b]{0.23\textwidth}
		\centering
		\includegraphics[width=\textwidth]{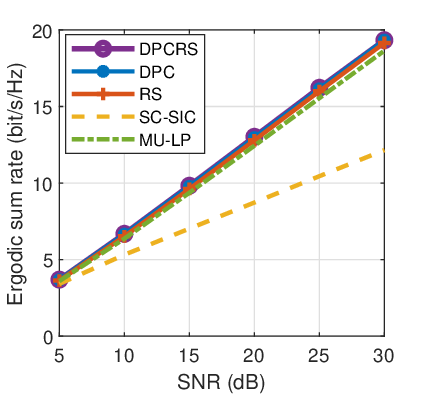}%
		\caption{$\alpha=0.9$}
	\end{subfigure}%
	\caption{Sum rate versus SNR comparison of different strategies with different partial CSIT inaccuracies, averaged over 100 random channel realizations, $K=2$, $N_t=4$, $\sigma_1^2=1, \sigma_2^2=1$. }
	\label{fig: bias1}
\end{figure}

\par
\begin{figure}
	\centering
	\begin{subfigure}[b]{0.23\textwidth}
		\centering
		\includegraphics[width=\textwidth]{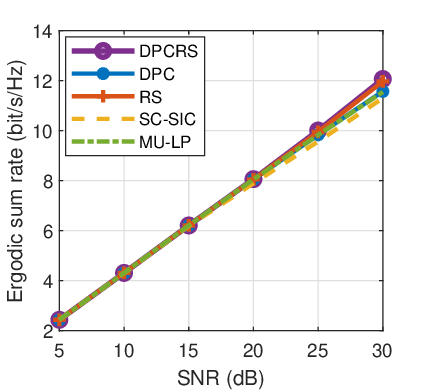}%
		\caption{$\alpha=0.3$}
	\end{subfigure}%
	~
	\begin{subfigure}[b]{0.23\textwidth}
		\centering
		\includegraphics[width=\textwidth]{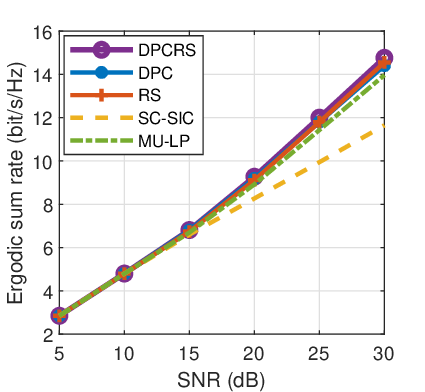}%
		\caption{$\alpha=0.6$}
	\end{subfigure}%
	~\\
	\begin{subfigure}[b]{0.23\textwidth}
		\centering
		\includegraphics[width=\textwidth]{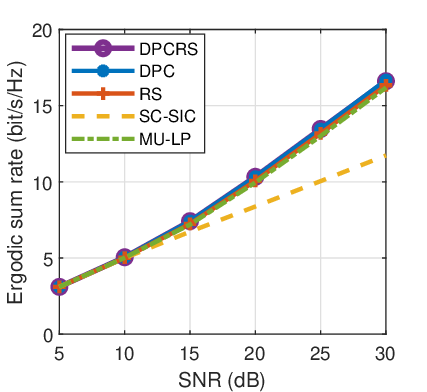}%
		\caption{$\alpha=0.9$}
	\end{subfigure}%
	\caption{Sum rate versus SNR comparison of different strategies with different partial CSIT inaccuracies, averaged over 100 random channel realizations, $K=2$, $N_t=4$, $\sigma_1^2=1, \sigma_2^2=0.09$. }
	\label{fig: bias03}
\end{figure}

 When $K=2$,  M-DPCRS, generalized RS and SC--SIC per group respectively reduces to 1-DPCRS,  1-layer RS and  SC--SIC. We use the term ``DPCRS" to represent both M-DPCRS and 1-DPCRS  and the term ``RS" to represent both generalized RS and 1-layer RS in the two-user case.  We first illustrate the system Ergodic Sum Rate (ESR) ($u_1=1, u_2=1$) versus SNR comparison  of different strategies considering diverse CSIT inaccuracies and channel strength disparities in Fig. \ref{fig: bias1} and  Fig. \ref{fig: bias03}. The individual QoS rate constraint of each user is set to 0, i.e., $R_k^{th}=0, \forall k\in\{1,2\}$. Without QoS rate constraint and unequal user weights, the WESR  problem for 1-layer RS reduces to the ESR maximization problem  studied in \cite{RS2016hamdi}. It has  been discovered in \cite{RS2016hamdi} the sum DoF achieved by solving the ESR  problem of 1-layer RS is
\begin{equation}
\label{eq: RS dof}
d_{\textrm{RS}}^{\star}=1+(K-1)\alpha,
\end{equation}
and $d_{\textrm{RS}}^{\star}$ has been shown to be the  sum DoF limit that could be achieved in MISO BC with partial CSIT. In comparison, the DoF achieved by optimally solving the ESR problem of MU--LP is 
$
d_{\textrm{MU--LP}}^{\star}=\max\{1,K\alpha\}.
$
As DoF captures the rate's asymptotic slope with respect to $\log_2(P_t)$, it is easy to identify the DoF according to the slope of the sum rate at high SNR from both Fig. \ref{fig: bias1} and  Fig. \ref{fig: bias03}. The values are calculated by scaling the high-SNR slopes to $\log_2(P_t)$. The DoF achieved by RS and DPCRS from both figures  are  close to the theoretically anticipated values calculated by (\ref{eq: RS dof}), i.e., $d_{\textrm{RS}}^{\star}=1.3,1,6, 1.9$ when $\alpha=0.3, 0.6, 0.9$, respectively. By using the same method, we obtain the DoF of MU--LP and DPC, which  also coincides with the theoretical DoF results, i.e.:  $d_{\textrm{MU--LP}}^{\star}=1,1.2, 1.8$ when $\alpha=0.3, 0.6, 0.9$, respectively. As the DoF of RS is  optimal, the DoF of DPCRS is the same as  RS. Though DPC is  capacity-achieving  in perfect CSIT, it is very sensitive to the CSIT inaccuracy. Its DoF is the same as MU--LP. As $\alpha$ decreases, the rate of DPC drops rapidly as MU--LP.
 In contrast, RS-assisted transmission strategies are more robust to the CSIT inaccuracy. Linearly precoded RS is not only more practical and low complex compared with DPC but also  achieves non-negligible rate gain over DPC when CSIT is imperfect. The rate gain of DPCRS and RS over all other strategies grows with SNR.  Specifically, both DPCRS and RS  achieves 16.12\% rate improvement over DPC and MU--LP when $\alpha=0.3$, SNR is 30 dB and users have equal channel variance as illustrated in Fig. \ref{fig: bias1}.  When there is a 10 dB average channel variance disparities as in Fig. \ref{fig: bias03}, the rate gain of RS decreases since there is a higher probability that the transmission reduces to single-user transmission by switching off the weaker user. In both figures, SC--SIC has the worst performance due to the fact that it does not exploit efficiently the spatial domain for interference management. The DoF of SC--SIC is  the worst and is limited to 1 (same as OMA) \cite{bruno2019wcl}. 

\par
\begin{figure}
	\begin{subfigure}[b]{0.23\textwidth}
		\centering
		\includegraphics[width=\textwidth]{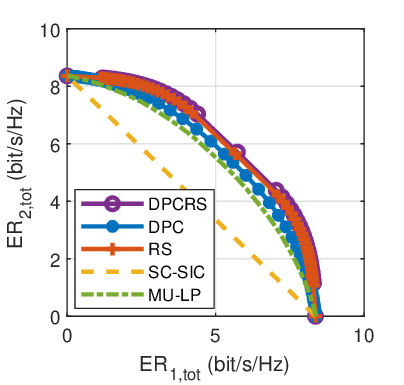}%
		\caption{$N_t=4, \sigma_2^2=1$}
	\end{subfigure}%
	~
	\begin{subfigure}[b]{0.23\textwidth}
		\centering
		\includegraphics[width=\textwidth]{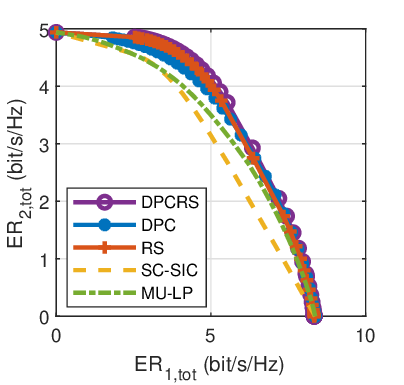}%
		\caption{$N_t=4, \sigma_2^2=0.09$}
	\end{subfigure}%
	~\\
	\begin{subfigure}[b]{0.23\textwidth}
		\centering
		\includegraphics[width=\textwidth]{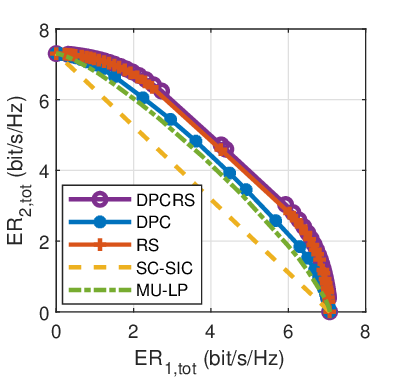}%
		\caption{$N_t=2, \sigma_2^2=1$}
	\end{subfigure}%
	~
	\begin{subfigure}[b]{0.23\textwidth}
		\centering
		\includegraphics[width=\textwidth]{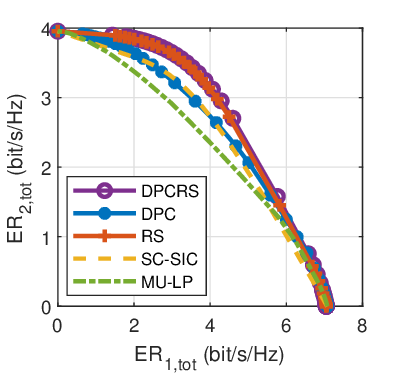}%
		\caption{$N_t=2, \sigma_2^2=0.09$}
	\end{subfigure}%
	\caption{Ergodic rate region comparison of different strategies with partial CSIT, averaged over 100 random channel realizations, SNR = 20 dB, $K=2$,  $\alpha=0.6$, $\sigma_1^2=1$. }
	\label{fig: Ergodic snr20 alpha06}
\end{figure}

 We further investigate the ER region in the two-user case.    Denote  the  optimized rate vector of problem (\ref{eq: DPC b}) and (\ref{eq: DPCRS b}) for the two users as  $\widehat{R}^{\textrm{x}\star}_{\pi, \mathbf{u}}=\{\widehat{R}_{\pi(1)}^{\textrm{x}\star}(\widehat{\mathbf{H}}),\widehat{R}_{\pi(2)}^{\textrm{x}\star}(\widehat{\mathbf{H}})\}$, where $\textrm{x}\in\{\textrm{``DPC", ``DPCRS"}\}$. The boundary points of the ER region $\mathcal{R}_{\pi}^{\textrm{x}}$ for a certain encoding order $\pi$ are first calculated over a set of different pairs of weights assigned to users and  averaging AR $\widehat{R}^{\textrm{x}\star}_{\pi, \mathbf{u}}$ over all channel uses for each pair of weights.
 The entire ER region of DPC/DPCRS  is the convex hull of the rate regions of all encoding orders, i.e, $\mathcal{R}_{\textrm{DPC}}=\textit{Conv}\left(\bigcup_{\pi}\mathcal{R}_{\pi}^{\textrm{x}}\right)$. Following \cite{RS2016hamdi}, the weight of user-1 is fixed as $u_1=1$ for each pair of user weights while the weight of user-2 changes as $u_2\in10^{[-3,-1,-0.95,\ldots,0.95,1,3]}$. As we study the largest rate region comparison, the   QoS rate constraint of each user is set to 0. Fig. \ref{fig: Ergodic snr20 alpha06} illustrates the rate region comparison of different strategies considering different number of transmit antennas and channel strength disparities. SNR is equal to 20 dB. In all subfigures, DPCRS achieves the largest rate region among all  strategies.  Comparing subfigure (a) and (c) in Fig. \ref{fig: Ergodic snr20 alpha06}, we observe that SC--SIC achieves the worst rate region. As there is no channel variance disparities among users, SC--SIC cannot properly utilize the power domain to manage interference.  As the number of transmit antennas decreases (from subfigure (a)/(b) to (c)/(d)), the rate region gap between DPCRS  and DPC/MU--LP becomes larger. Each user in DPC and MU--LP directly decodes its intended stream, and the   pressure of interference management is at the transmitter.  
In comparison, both DPCRS and RS utilize the common stream to enable each user the capability of partially decoding the interference and partially treating the interference as noise. Both of them are  more robust to various number of transmit antennas and user deployments.

\begin{figure}
	\begin{subfigure}[b]{0.23\textwidth}
		\centering
		\includegraphics[width=\textwidth]{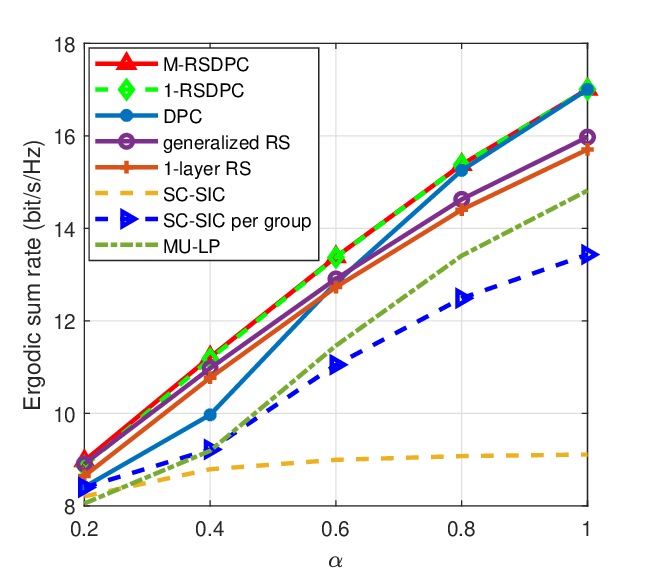}
		\caption{$\mathbf{r}_k^{th}=[0.1,\ldots,0.5]$ bit/s/Hz, $N_t=4$, $\sigma_1^2=\sigma_2^2=\sigma_3^2=1$.}
	\end{subfigure}%
	~ 
	\begin{subfigure}[b]{0.23\textwidth}
		\centering
		\includegraphics[width=\textwidth]{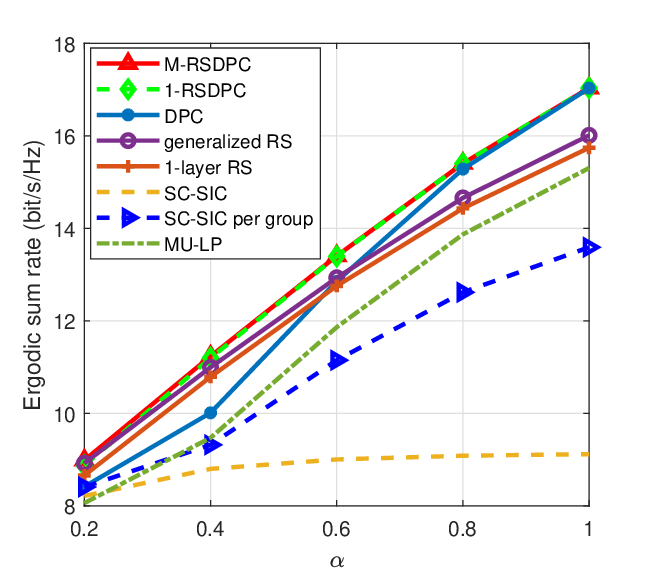}
		\caption{$R_k^{th}=0$,  $N_t=4$, $\sigma_1^2=\sigma_2^2=\sigma_3^2=1$.}
	\end{subfigure}%
	~ \\
	\begin{subfigure}[b]{0.23\textwidth}
		\centering
		\includegraphics[width=\textwidth]{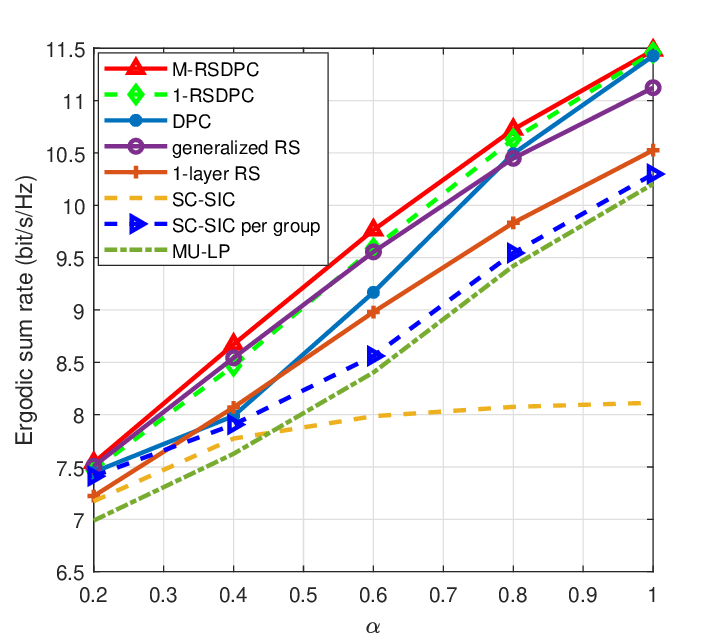}
		\caption{$R_k^{th}=0$,  $N_t=2$, $\sigma_1^2=\sigma_2^2=\sigma_3^2=1$.}
	\end{subfigure}%
	~ 
	\begin{subfigure}[b]{0.23\textwidth}
		\centering
		\includegraphics[width=\textwidth]{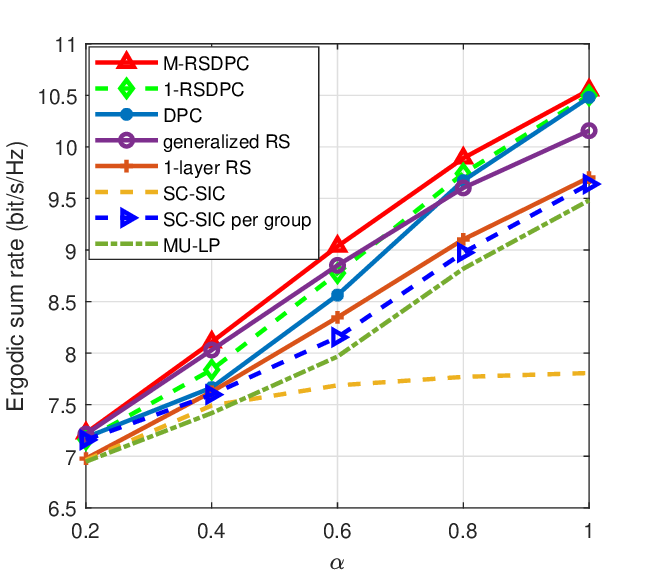}%
		\caption{$R_k^{th}=0$,  $N_t=2$, $\sigma_1^2=\sigma_2^2=1,\sigma_3^2=0.3$.}
	\end{subfigure}
	\caption{Ergodic sum rate versus CSIT inaccuracy $\alpha$  comparison of different strategies, averaged over 100 random channel realizations, $K=3$,  SNR = 20 dB.}
	\label{fig: SRwithAlpha bias11}
\end{figure}

\subsection{Three-user case}
\par When $K=3$,  the ESR of all the eight strategies versus CSIT inaccuracy are compared in Fig. \ref{fig: SRwithAlpha bias11} with different QoS rate constraints, network loads and user deployments. SNR is 20 dB. In subfigure (a), the individual QoS rate constraint increases with CSIT accuracy.  For $\alpha=[0.2,0.4,0.6,0.8,1]$, the corresponding rate constraint for  user-$k$ ($k\in\{1,2,3\}$) changes as $\mathbf{r}_k^{th}=[0.1,0.2,0.3,0.4,0.5]$ bit/s/Hz. 
In all subfigures, the ESR of DPC, MU--LP, and SC--SIC-assisted strategies decrease dramatically as $\alpha$ decreases from 1 to 0.2  due to the drop-off of CSIT accuracy. The ESR gap between M-DPCRS/1-DPCRS and DPC is more obvious in the region with strong CSIT inaccuracy. Thanks to their ability to partially decode interference and partially treat interference as noise, all RS-assisted transmission strategies  are more robust to the CSIT inaccuracy.   In Fig. \ref{fig: SRwithAlpha bias11}(a) and Fig. \ref{fig: SRwithAlpha bias11}(b) with underloaded network loads, generalized RS and 1-layer RS, using linear precoder  for all streams,  achieve explicit ESR improvement over DPC when $\alpha$ is less than 0.6. This observation further confirms the  powerful interference management capability of RS in the multi-antenna BC. In Fig. \ref{fig: SRwithAlpha bias11}(c) and Fig. \ref{fig: SRwithAlpha bias11}(d) where network loads are overloaded and users suffer from stronger inter-user interference, we observe that M-DPCRS (generalized RS) has higher rate  than 1-DPCRS (1-layer RS). By increasing the number of layers of common streams in RS, inter-user interference is better managed and ESR is further improved even though there is no DoF increase. 

\section{Conclusion}
\label{sec: conclusion}
\par To conclude, we propose a novel strategy, namely, DPCRS  in this work  for assessing the  rate region of multi-antenna BC with partial CSIT  by incorporating RS with DPC.
By splitting the user messages at the transmitter into common and private parts, and use DPC to encode the private parts, DPCRS not only enables the ability to partially decode the interference and partially treat interference as noise, but also further restrains the multi-user interference among private messages. Numerical results   first show that the existing linearly precoded RS,  benefiting from its robustness in partial CSIT,  outperforms DPC if CSIT is sufficiently inaccurate in MISO BC. Most importantly, the  rate region improvement of linearly precoded RS over DPC comes with much lower hardware and computational complexities.   This is sharply different from the observations in perfect CSIT where DPC outperforms all linearly precoded strategies. Moreover, we show that the proposed DPCRS not only  enlarges the  rate region of  MISO BC with partial CSIT but  is  more robust to CSIT inaccuracies, network loads and user deployments compared with DPC and other existing transmission strategies. 
 
\section*{Acknowledgement}
\par The authors are deeply indebted to Dr. Hamdi Joudeh for his useful  insights and  suggestions.

\section*{Appendix}
\par We follow  Theorem 1 in \cite{yang2005impact} and obtain the following Lower Bound (LB) of the achievable rates of DPC and DPCRS, which is given as:
\begin{equation}
\label{eq: LB DPC}
\begin{aligned}
&R_{\pi(1)}^{\textrm{x}}=\log_2\left(1+\frac{|\mathbf{h}_{\pi(1)}^H\mathbf{p}_{\pi(1)}|^2}{|\mathbf{h}_{\pi(1)}^H\mathbf{p}_{\pi(2)}|^2+1}\right), \\
&R_{{\pi(2)}}^{\textrm{x}}=
\log_2\left(1+\frac{|\mathbf{h}_{\pi(2)}^H\mathbf{p}_{\pi(2)}|^2}{\sigma_{e,2}^2\|\mathbf{p}_{\pi(1)}\|^2+1
}\right), 
\end{aligned}
\end{equation}
where $\textrm{x}\in\{\textrm{``DPC'', ``DPCRS"}\}$.
 The procedure of deriving (\ref{eq: LB DPC}) follows \cite{yang2005impact}. 
$R_{\pi(1)}^{\textrm{x}}$ is trivial. For user-$\pi(2)$, its corresponding received signal is:
\begin{equation}
\resizebox{.48\textwidth}{!} {$\begin{aligned}
&y_{\pi(2)}=\mathbf{h}_{\pi(2)}^H\mathbf{p}_{\pi(1)}s_{\pi(1)}+\mathbf{h}_{\pi(2)}^H\mathbf{p}_{\pi(2)}s_{\pi(2)}+n_{\pi(2)}\\
&=\mathbf{h}_{\pi(2)}^H\mathbf{p}_{\pi(2)}s_{\pi(2)}+\widehat{\mathbf{h}}_{\pi(2)}^H\mathbf{p}_{\pi(1)}s_{\pi(1)}+\widetilde{\mathbf{h}}_{\pi(2)}^H\mathbf{p}_{\pi(1)}s_{\pi(1)}+n_{\pi(2)}\\
&=h_{\pi(2)}s_{\pi(2)}+\widehat{h}_{\pi(2)}s_{\pi(1)}-w_{\pi(2)}s_{\pi(1)}+n_{\pi(2)},
\end{aligned}$}
\end{equation}
where $h_{\pi(2)}=\mathbf{h}_{\pi(2)}^H\mathbf{p}_{\pi(2)}$, $\widehat{h}_{\pi(2)}=\widehat{\mathbf{h}}_{\pi(2)}^H\mathbf{p}_{\pi(1)}$ and $w_{\pi(2)}=-\widetilde{\mathbf{h}}_{\pi(2)}^H\mathbf{p}_{\pi(1)}$ and we have $\gamma_k=|h_{k}|^2=|\mathbf{h}_{k}^H\mathbf{p}_{k}|^2$ and $\mathbb{E}\{|w_{\pi(2)}|^2\}\leq \sigma_{e,k}^2\|\mathbf{p}_{\pi(1)}\|^2$. Following Lemma 1 of \cite{yang2005impact}, we  obtain that  LB (\ref{eq: LB DPC}) is achievable. 

\par To compare RS with DPC and DPCRS,  we follow the same method and obtain the LB of the  achievable rate of RS, which is given as:
\begin{equation}
\label{eq: LB RS}
\begin{aligned}
R_{\pi(1)}^{\textrm{RS}}&=R_{\pi(1)}^{\textrm{x}},\\ 
R_{{\pi(2)}}^{\textrm{RS}}&=
\log_2\left(1+\frac{|\mathbf{h}_{\pi(2)}^H\mathbf{p}_{\pi(2)}|^2}{|\widehat{\mathbf{h}}_{\pi(2)}^H\mathbf{p}_{\pi(1)}|^2+\sigma_{e,2}^2\|\mathbf{p}_{\pi(1)}\|^2+1
}\right).
\end{aligned}
\end{equation}
Note that the encoding order $\pi$ specified in equation (\ref{eq: LB RS})  decides which user to approximate the rate to its LB.

\par
\begin{figure} [t!]
	\centering
	\begin{subfigure}[b]{0.23\textwidth}
		\centering
		\includegraphics[width=\textwidth]{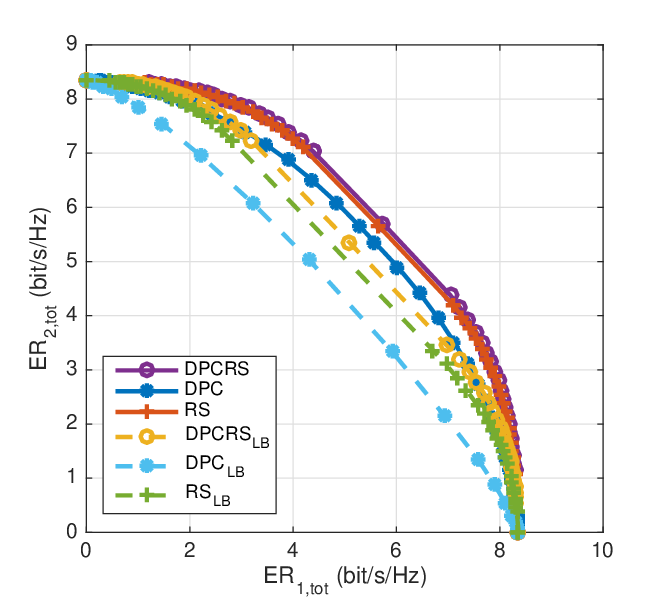}%
		\caption{$N_t=4, \sigma_2^2=1$}
	\end{subfigure}%
	~
	\begin{subfigure}[b]{0.23\textwidth}
		\centering
		\includegraphics[width=\textwidth]{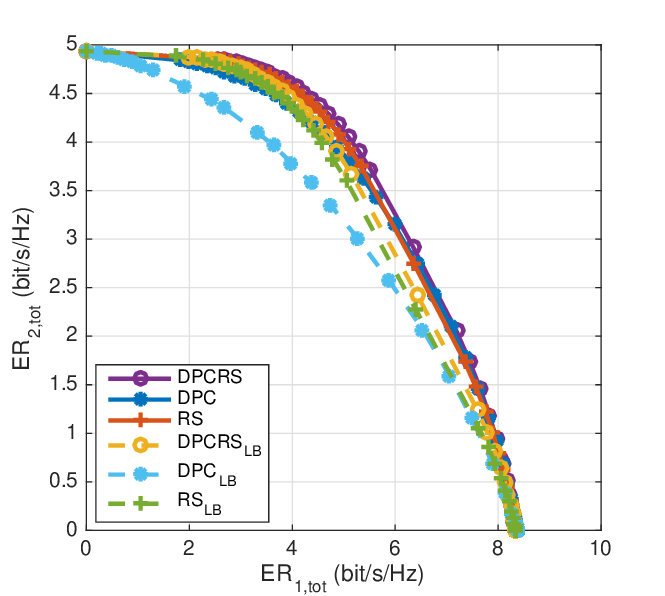}%
		\caption{$N_t=4, \sigma_2^2=0.09$}
	\end{subfigure}%
	~\\
	\begin{subfigure}[b]{0.23\textwidth}
		\centering
		\includegraphics[width=\textwidth]{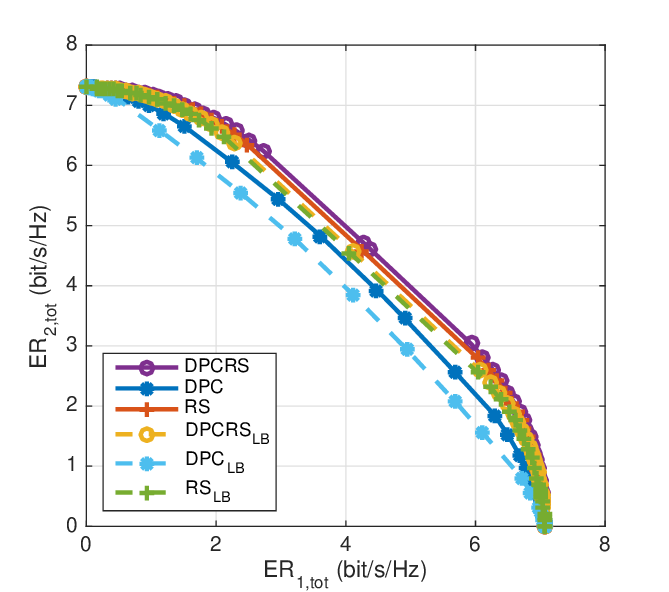}%
		\caption{$N_t=2, \sigma_2^2=1$}
	\end{subfigure}%
	~
	\begin{subfigure}[b]{0.23\textwidth}
		\centering
		\includegraphics[width=\textwidth]{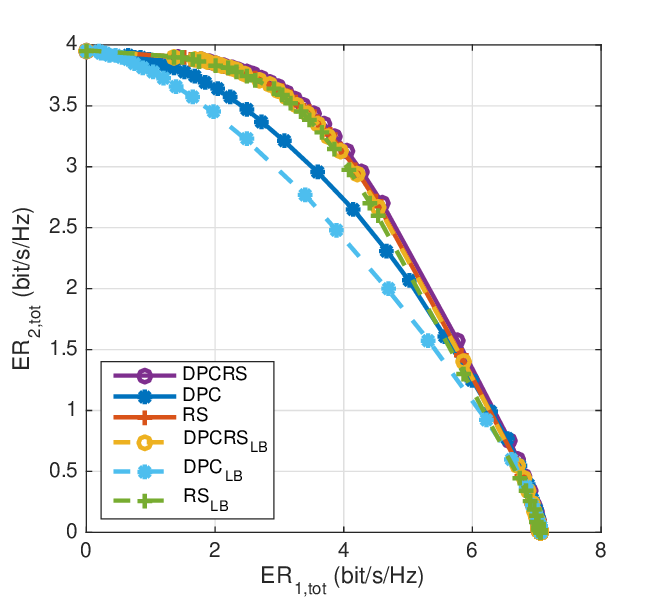}%
		\caption{$N_t=2, \sigma_2^2=0.09$}
	\end{subfigure}%
	\caption{Ergodic rate region comparison of different strategies and their corresponding lower bounds with partial CSIT, averaged over 100 random channel realizations, SNR = 20 dB, $K=2$,  $\alpha=0.6$, $\sigma_1^2=1$. }
	\label{fig: Ergodic snr20 alpha06 LB}
\end{figure}

 We follow the method adopted in Section \ref{sec: simulation two user}  to illustrate the Ergodic Rate  (ER) regions of  DPC, DPCRS, RS as well as their corresponding LBs  based on (\ref{eq: LB DPC})  and (\ref{eq: LB RS}) in Fig. \ref{fig: Ergodic snr20 alpha06 LB}. The ER regions of LBs of DPC, DPCRS, RS are denoted by ``$\textrm{DPC}_\textrm{LB}$", ``$\textrm{DPCRS}_\textrm{LB}$", ``$\textrm{RS}_\textrm{LB}$", respectively.
We first observe from Fig. \ref{fig: Ergodic snr20 alpha06 LB} that the ER region obtained based on the LB (\ref{eq: LB DPC}) is indeed smaller than that obtained based on (\ref{eq: rate DPC}) and (\ref{eq: DPC private 1-RS}). We also  find from Fig. \ref{fig: Ergodic snr20 alpha06 LB}  that the ER region of the LB of DPCRS is always larger than that of RS or DPC. RS achieves an obvious LB improvement over DPC and it is closer to the LB of DPCRS. Moreover, when $N_t=2$, the ER region of the LB of RS is even larger than  the ER region  of DPC (which might be the outer bound of DPC), but with a much lower  complexity. Therefore, we draw the conclusion that RS strategies with linear or non-linear precoding are promising strategies for future wireless communication networks since they outperform conventional DPC and are able to come with lower complexity.

\bibliographystyle{IEEEtran}
\bibliography{reference}
\end{document}